\documentclass[12pt]{JHEP3}

\usepackage{epsfig}

\newcommand{\newc}{\newcommand}
\def\l{\lambda}
\def\p{\phi}

\def\mg{m_{3/2}}

\newc{\vev}[1]{\langle #1 \rangle}

\newc{\eq}[1]{\mbox{(\ref{eq:#1})}}

\title{The Peccei--Quinn Field as Curvaton}

\author{Konstantinos Dimopoulos,$^{ab}$
George Lazarides,$^c$ David H. Lyth$^a$ and
Roberto Ruiz de Austri$^c$ \\
$^a$Department of Physics, Lancaster
University, Lancaster LA1 4YB, England \\
$^b$Department of Physics, University of
Oxford, 1 Keble Road, Oxford OX1 3NP,
England \\
$^c$Physics Division, School of Technology,
Aristotle University of Thessaloniki, \\
Thessaloniki 54124, Greece  \\
E-mail:
\email{konst.dimopoulos@lancaster.ac.uk},
\email{lazaride@eng.auth.gr},
\email{d.lyth@lancaster.ac.uk},
\email{rruiz@gen.auth.gr}}

\preprint{UT-STPD-1/03}

\abstract{A simple extension of
the minimal supersymmetric
standard model which naturally and
simultaneously solves the strong
CP and $\mu$ problems via a
Peccei-Quinn and a continuous R
symmetry is considered. This model
is supplemented with hybrid
inflation and leptogenesis, but
without taking the specific
details of these scenarios. It is
shown that the Peccei-Quinn field
can successfully act as a curvaton
generating the total curvature
perturbation in the universe in
accord with the cosmic background
explorer measurements. A crucial
phenomenon, which assists us to
achieve this, is the `tachyonic
amplification' of the perturbation
acquired by this field during
inflation if the field, in its
subsequent evolution, happens to
be stabilized for a while near a
maximum of the potential. In this
case, the contribution of the
field to the total energy density
is also enhanced (`tachyonic
effect'), which helps too. The
cold dark matter in the universe
consists, in this model, mainly of
axions which carry an isocurvature
perturbation uncorrelated with the
total curvature perturbation.
There are also lightest sparticles
(neutralinos) which, like the
baryons, originate from the
inflationary reheating and, thus,
acquire an isocurvature
perturbation fully correlated with
the curvature perturbation. So, the
overall isocurvature perturbation
has a mixed correlation with the
adiabatic one. It is shown that the
presently available bound on such
an isocurvature perturbation from
cosmic microwave background
radiation and other data is
satisfied. Also, the
constraint on the non-Gaussianity of
the curvature perturbation obtained
from the recent Wilkinson microwave
anisotropy probe data is fulfilled
thanks to the `tachyonic effect'.}


\begin{document}

\section{Introduction}
\label{sect:introduction}

\par
The recent data \cite{wmap}
on the acoustic peaks of the angular
power spectrum of the cosmic microwave
background radiation (CMBR) strongly
favor the idea of inflation, which
offers the most elegant solution to the
outstanding problems of standard big
bang cosmology and exterminates unwanted
relics such as monopoles. Moreover,
inflation is now generally accepted as
the most likely origin of the primordial
density perturbations which are needed
for explaining the structure formation
in the universe \cite{llbook}. The usual
assumption is \cite{llbook,lectures}
that these perturbations come solely
from the slowly rolling inflaton field.
In this case, the observed density
perturbations are expected to be purely
adiabatic, since fluctuations in the
inflaton cannot cause an isocurvature
perturbation. Also, significant
non-Gaussianity is not encountered
\cite{onefield} in the usual one-field
inflationary models, while the
non-Gaussianity which is possible
\cite{multifield} in multifield models
requires extreme fine-tuning of the
initial conditions. However, although
adiabatic and Gaussian perturbations
are perfectly consistent with the
present data, appreciable
non-Gaussianity \cite{nongauss} or the
presence of a significant isocurvature
density perturbation \cite{trd,agws}
cannot be excluded by observations.

\par
An alternative possibility \cite{curv},
which has recently attracted attention
\cite{curv1,luw02,curv2,dimo,curv3,gl,
curv4,prep}, is that the adiabatic
density perturbations originate from the
vacuum perturbations during inflation of
some light `curvaton' field different
from the inflaton. In the curvaton
scenario, significant non-Gaussianity
may easily appear because the curvaton
density is proportional to the square of
the curvaton field amplitude. Also, the
curvaton density perturbations can lead,
after curvaton decay, to isocurvature
perturbations in the densities of the
various components of the cosmic fluid.
In the simplest case, the residual
isocurvature perturbations are either
fully correlated or fully
anti-correlated with the adiabatic
density perturbation, with a calculable
and generally significant relative
magnitude. In the presence of axions,
however, the correlation is, in general,
mixed. It is important to be noted that
the curvaton hypothesis makes
\cite{dimo} the task of constructing
viable models of inflation much easier,
since it liberates us from the very
restrictive requirement that the
inflaton is responsible for the
curvature perturbations.

\par
The most compelling extension of the
standard model (SM) of particle physics
is its supersymmetric (SUSY) version,
the so-called minimal supersymmetric
standard model (MSSM). It is, however,
clear that even the MSSM must be part
of a larger scheme since it leaves a
number of fundamental questions
unanswered. For instance, the strong CP
and $\mu$ problems cannot be solved
in MSSM. Also, it is not easy to
generate in MSSM the observed baryon
asymmetry of the universe by
electroweak sphaleron processes or to
realize inflation.

\par
The strong CP problem can be elegantly
resolved by including a Peccei-Quinn (PQ)
symmetry \cite{pq77} broken spontaneously
at an intermediate mass scale, which can
be easily generated within the supergravity
(SUGRA) extension of MSSM. The PQ field,
which breaks the PQ symmetry, corresponds,
in this case, to a flat direction in field
space lifted by non-renormalizable
interactions. Moreover, the $\mu$ parameter
of MSSM can be generated \cite{kn84} from
the PQ scale. A minimal extension
of MSSM, which solves the strong CP and
$\mu$ problems along these lines, has been
constructed in Ref.~\cite{ls98}. A key
ingredient was a global $U(1)$ R symmetry
obeyed by the superpotential. This model
has been further extended \cite{inf} to
simple SUSY grand unified theory (GUT)
models leading to hybrid inflation
\cite{hybrid} and yielding successful
baryogenesis via a primordial leptogenesis
\cite{lepto,leptoinf} in accord with the
data on neutrino masses and mixing. Also,
the R symmetry implies exact baryon number
conservation, thereby explaining proton
stability.

\par
In this paper, we examine whether, in the
above models, the PQ field, which possesses
an almost flat potential, can also play the
role of the curvaton generating the
adiabatic curvature perturbations with a
possibly significant non-Gaussian component,
while accompanied also by an isocurvature
density perturbation. An important
requirement is that the PQ field is
effectively massless during inflation. In
view of the
quasi-flatness of the potential, one would
imagine that this condition could be easily
fulfilled by just ensuring that the value
of the PQ field during inflation is not too
large. However, as it is well-known
\cite{crisis,cllsw}, large SUGRA
corrections during inflation can destroy
the quasi-flatness of the PQ potential and,
thus, invalidate the possibility of using
the PQ field as curvaton. Therefore, it is
crucial to assume that some mechanism
\cite{noscale} is employed to keep these
corrections under control.

\par
After the end of inflation, the inflaton
starts performing coherent damped oscillations
and eventually decays reheating the universe.
We assume that the SUSY breaking corrections
to the PQ potential which arise \cite{drt95}
from the finite energy density of the
oscillating inflaton field are also relatively
small. Under these circumstances, the PQ field
remains overdamped and, thus, undergoes a slow
evolution for quite some time after the
termination of inflation. At later cosmic
times, however, it settles into coherent damped
harmonic oscillations about zero or the PQ
vacua, depending on its value at the end of
inflation. Actually, we find that the values
of the PQ field at the end of inflation fall
into well-defined alternating bands which lead
to the trivial (false) vacuum or the PQ vacua
in turn. Of course, bands leading to the
trivial vacuum must be excluded.

\par
The perturbation acquired by the (light) PQ
field during inflation evolves at subsequent
times and, when the system settles into
damped quadratic oscillations about the PQ
vacua, yields a stable perturbation in the
energy density of the oscillating field.
After the PQ field decays, this perturbation
is transferred to the radiation dominated
plasma, thereby
generating the total curvature perturbation.
We find that, if the PQ field, as it evolves,
happens to stay near a maximum of the
potential for some time, its primordial
perturbation from inflation is drastically
enhanced and, thus, the resulting energy
density perturbation of the oscillating
field can be quite sizeable. Also, the
contribution of the oscillating field to
the total energy density comes out, in this
case, larger due to the delay in the onset
of the field oscillations caused by the
temporary rest of the field near a maximum.
Thus, the final
curvature perturbation can be adequate to
meet the cosmic background explorer (COBE)
results \cite{cobe}. This `tachyonic
amplification' of the inflationary
perturbation of the PQ field as well as the
enhancement of its contribution to the
energy density, which occur well after the
end of inflation, are crucial
for the viability of our scheme.

\par
The cold dark matter (CDM) of the universe,
in our case, consists of axions and
lightest neutralinos. The latter originate
from the late decay of the gravitinos which
are generated during the reheating process
which follows inflation. In accordance to
the curvaton hypothesis, the inflaton and,
thus, the radiation which emerges from its
decay at reheating possess practically no
partial curvature perturbation. Therefore,
the baryons, which originate from a
primordial leptogenesis taking place at
reheating, as well as the neutralinos
carry negligible partial curvature
perturbation. Then, at curvaton decay,
where its curvature perturbation is
transferred to the radiation, the baryons
and the neutralinos acquire an isocurvature
perturbation, which is fully correlated
with the total curvature perturbation. The
axions, which come into play at the QCD
phase transition, also acquire
\cite{axionisoc} an isocurvature
perturbation which is, though, uncorrelated
with the total curvature perturbation. So,
we finally obtain an isocurvature
perturbation of mixed correlation with the
curvature perturbation. We apply the most
stringent available restriction on such
isocurvature perturbations. This restriction
results from the set of the CMBR and other
data used in Ref.~\cite{gl}.

\par
In our scheme, the curvaton decays somewhat
before dominating the energy density of the
universe. This can generate appreciable
non-Gaussianity in the curvature perturbation.
We consider the bound \cite{nongauss} on the
possible non-Gaussian component of the
curvature perturbation from the recent CMBR
data which were obtained by the Wilkinson
microwave anisotropy probe (WMAP) satellite.
The `tachyonic effect', i.e. the possible
rest of the field near a maximum of the
potential for a period of time, is
essential also for the fulfillment of the
non-Gaussianity requirement.

\par
The paper is organized as follows. In
section~\ref{sect:solution}, we summarize the
minimal extension of MSSM which solves the
strong CP and $\mu$ problems via a PQ and a R
symmetry. We also indicate the salient features
of the simple SUSY GUTs which further extend
this model and yield hybrid inflation and
successful baryogenesis. In
section~\ref{sect:evolution}, we outline the
evolution of the PQ system during inflation,
the subsequent inflaton oscillations and the
period after
reheating. In section~\ref{sect:curvaton}, we
discuss the conditions under which the PQ
field can successfully act as a curvaton. In
particular, we consider the constraints on the
total curvature and isocurvature perturbation
from the COBE results and other data.
The requirements from the CDM in the universe
are also taken into account.
Section~\ref{sect:analysis} is devoted to the
detailed numerical analysis of the evolution
of the PQ field and its perturbations which
originate from inflation. The various
requirements of section~\ref{sect:curvaton}
and the constraints from the non-Gaussianity
of the curvature perturbation are numerically
studied. Finally, in
section~\ref{sect:conclusion}, we summarize
our main conclusions.

\section{A simultaneous solution of the strong
CP and $\mu$ problems}
\label{sect:solution}

\par
It is well-known that, in the context of the
SUGRA extension of MSSM, appropriate flat
directions in field space can generate an
intermediate scale
\begin{equation}
M_I\sim (\mg m_{P})^{\frac{1}{2}}\sim
10^{11}-10^{12}\,{\rm GeV},
\end{equation}
where $\mg\sim 1\,{\rm TeV}$ is the SUSY
breaking scale and $m_P\equiv
(8\pi G_N)^{-1/2}\simeq 2.44\times 10^{18}~
{\rm GeV}$ is the `reduced' Planck mass with
$G_N$ being the Newton's constant. It is
then natural to identify the scale $M_I$
with the breaking scale $f_a$ of a
PQ symmetry $U(1)_{\rm PQ}$ \cite{pq77},
which resolves the strong CP problem.
Moreover, the $\mu$ parameter of MSSM could
be obtained as $\mu\sim f_a^2/m_P\sim\mg$
\cite{kn84}.

\par
This idea has been elegantly implemented in
Ref.~\cite{ls98} in the presence of a
non-anomalous R symmetry $U(1)_R$. To this
end, an extra pair of gauge singlet chiral
superfields $N$, $\bar{N}$ with the
superpotential couplings
\begin{equation}
\delta W=\frac{\l}{2m_P}N^2\bar{N}^2+
\frac{\beta}{m_P}N^2h_1h_2
\label{eq:superpotential}
\end{equation}
was added in MSSM. Here, $h_1$ and $h_2$ are
the two Higgs doublets coupling to the down
and up type quarks respectively, and $\l$,
$\beta$ are dimensionless parameters,
which can be made positive by appropriate
field redefinitions. The PQ and R charges
are
\begin{eqnarray}
PQ:~N(-1),~\bar{N}(1),~h_1(1),~h_2(1),
\nonumber \\
R:~N(1),~\bar{N}(0),~h_1(0),~h_2(0)
\end{eqnarray}
with the `matter' (quark and lepton)
superfields having $PQ=-1/2$, $R=1$.
The superpotential has $R=2$. Note that
global continuous symmetries such as
our PQ and R symmetry can effectively
arise \cite{continuous} from the rich
discrete symmetry groups encountered in
many compactified string theories (see
e.g. Ref.~\cite{discrete}). It should
be mentioned in passing that such
discrete symmetries can also provide
\cite{doublet} an alternative solution
to the $\mu$ problem.

\par
The part of the tree-level scalar potential
which is relevant for the PQ symmetry
breaking is derived from the superpotential
term $\lambda N^2\bar{N}^2/2m_P$ and, after
soft SUSY breaking mediated by minimal SUGRA,
is given by \cite{ls98}
\begin{equation}
V_{\rm PQ}=\left(\mg^2+\l^2\left|\frac
{N\bar{N}}{m_P}\right|^2\right)(|N|^2+
|\bar{N}|^2)+\left(A\mg\l\frac{N^2\bar{N}^2}
{2m_P}+{\rm h.c.}\right),
\label{eq:vpq}
\end{equation}
where $A$ is the dimensionless coefficient
of the soft SUSY breaking term
corresponding to the superpotential term
$\lambda N^2\bar{N}^2/2m_P$.

\par
The potential $V_{\rm PQ}$ is minimized along
the field direction where $N$, $\bar{N}$ have
the same magnitude and their phases are
arranged so that the last term in the right
hand side (RHS) of Eq.~(\ref{eq:vpq}) becomes
negative. The potential then takes the form
\cite{ls98}
\begin{equation}
V_{\rm PQ}=2|N|^2\mg^2\left(1-|A|\l
\frac{|N|^2}{2\mg m_P}+\l^2\frac{|N|^4}
{\mg^2m_P^2}\right).
\label{eq:N-pot}
\end{equation}
For $|A|>4$, the absolute minimum of the
potential is at
\begin{equation}
|\vev{N}|=|\vev{\bar{N}}|=
\left(\frac{|A|+(|A|^2-12)^{\frac{1}{2}}}
{6\l}\right)^{\frac{1}{2}}(\mg m_P)^{
\frac{1}{2}}\sim (\mg m_P)^{\frac{1}{2}}.
\end{equation}
The vacuum expectation values (VEVs) of $N$,
$\bar{N}$ break spontaneously the symmetry
$U(1)_{\rm PQ}\times U(1)_R$ down to the
discrete `matter parity' symmetry
$Z^{\rm mp}_2$ under which the `matter'
superfields change sign. Substituting these
VEVs in Eq.~\eq{superpotential}, we see that
a $\mu$-term with $\mu\sim\mg$ is generated.

\par
This scheme has been embedded \cite{inf} in
concrete SUSY GUT models which naturally
lead to hybrid inflation \cite{hybrid}
followed by a reheating process which
satisfies the gravitino constraint \cite{ekn}.
These models also yield successful
baryogenesis via a primordial leptogenesis
\cite{lepto} which takes place \cite{leptoinf}
non-thermally at reheating by the decay
of right handed neutrino superfields
which emerge from the inflaton decay. The
observed baryon asymmetry of the universe
can be achieved in accord with the present
neutrino oscillation data.

\par
In these models, the PQ and R charges of $N$,
$\bar{N}$ and the MSSM superfields are the
same as above with the exception of the PQ
charge of the left (right) handed quark and
lepton superfields which is now $PQ=-1$
($PQ=0$). This fact implies that the discrete
`matter parity'  symmetry ($Z^{\rm mp}_2$) is
now not contained in $U(1)_{\rm PQ}$ and has,
thus, to be independently imposed. The VEVs
of $N$, $\bar{N}$ break $U(1)_{\rm PQ}\times
U(1)_R$ completely, while $Z^{\rm mp}_2$ again
remains exact.

\par
A closer look reveals that the QCD instantons
break $U(1)_{\rm PQ}$ explicitly to its $Z_6$
subgroup, which is then broken by $\vev{N}$,
$\vev{\bar{N}}$ (as in Refs.~\cite{jean,talks}).
This would lead \cite{sikivie} to a disastrous
domain wall production if the PQ transition
occurred after inflation. Avoidance of this
disaster would then require extending
\cite{walls} the model so that such
spontaneously broken discrete symmetries do
not appear. The soft SUSY breaking terms
break $U(1)_R$ explicitly to its $Z_2$
subgroup which, in the presence of
$Z^{\rm mp}_2$, is equivalent to the
$Z_2$ under which only $N$ changes sign
(compare with Refs.~\cite{jean,talks}). The
breaking of this $Z_2$ by $\vev{N}$ would
also be catastrophic if it took place after
inflation. Fortunately, as we will argue
in section~\ref{sect:evolution}, the $N$,
$\bar{N}$ fields emerge with non-zero
(homogeneous) values at the end of inflation
and, thus, the above discrete symmetries are
broken. Also, there is no PQ transition
after inflation. Moreover, as we will
explain in section~\ref{subsect:band}, the
system is led to the same minimum of the PQ
potential everywhere in space. Consequently,
domain walls do not form and there is no
need to further complicate the model.

\par
The fact that $N,~\bar{N}\neq 0$ after
inflation solves yet another problem. It
has been shown \cite{jean,talks} that the
trivial local minimum of $V_{\rm PQ}$ at
$N=\bar{N}=0$ is separated from the PQ
(absolute) minimum by a sizeable
potential barrier. This picture holds
\cite{jean,talks} at all temperatures
below the reheat temperature
$T_{\rm reh}\lesssim 10^9~{\rm GeV}$
\cite{ekn} even if the
one-loop temperature corrections to the
potential are included. Thus, after
reheating, a PQ transition would have been
impossible if the system was in the
trivial vacuum. Such a transition could
have been even more impossible during
inflaton oscillations if the
system emerged after inflation being in
the trivial vacuum. This is due to the
fact that the PQ field typically acquires
\cite{drt95} a positive ${\rm mass}^2$ of
order $H^2$ ($H$ is the Hubble parameter)
via its coupling to the inflaton (see
section~\ref{sect:evolution}).

\section{The cosmological evolution of the
PQ field}
\label{sect:evolution}

\par
We restrict ourselves on the field direction
$\bar{N}^*=\pm N\exp{[i(\alpha+\pi)/2]}$
($\alpha$ is the phase of $A$ in
Eq.~\eq{vpq}) along which $V_{\rm PQ}$ is
minimized and takes the form in
Eq.~(\ref{eq:N-pot}). Rotating $N$ on the
real axis by an appropriate R transformation,
we can then define the canonically
normalized real scalar PQ field $\p=2N$,
which breaks the PQ symmetry and generates
the $\mu$-term as discussed in
section~\ref{sect:solution}. Eq.~\eq{N-pot}
then becomes
\begin{equation}
V_{\rm PQ}=\frac{1}{2}\mg^2\p^2\left(1-|A|\l
\frac{\p^2}{8\mg m_P}+\l^2\frac{\p^4}{16\mg^2
m_P^2}\right).
\label{eq:PQ-pot}
\end{equation}
For $|A|>4$, this potential has a local
minimum at $\p=0$ and absolute minima at
\begin{equation}
\vev{\p}^2\equiv f_a^2=\frac{2}{3\l}\left(|A|+
\sqrt{|A|^2-12}\right)\mg m_P
\label{eq:fa}
\end{equation}
with $f_a>0$. We can shift the potential
$V_{\rm PQ}$ in Eq.~\eq{PQ-pot} by adding to
it the constant
\begin{equation}
V_0=\frac{1}{108\l}\left(|A|+
\sqrt{|A|^2-12}\right)\left[|A|\left(|A|+
\sqrt{|A|^2-12}\right)-24\right]\mg^3 m_P,
\label{eq:v0}
\end{equation}
so that it vanishes at its absolute minima.

\par
In order to be able to study the
cosmological evolution of $\p$, we should
first include the SUSY breaking effects
which arise \cite{crisis,cllsw,drt95} from
the fact that the energy density in the
early universe is finite. During inflation
and the subsequent inflaton oscillations,
SUSY breaking is transmitted to the PQ
system via its coupling to the inflaton
given by the SUGRA scalar potential. The
scale of the resulting SUSY breaking
terms is set by the Hubble parameter $H$
and, thus, these terms dominate over the
terms from hidden sector SUSY breaking
as long as $H\gtrsim m_{3/2}$.

\par
We should point out that, during the
era of inflaton oscillations, the PQ
potential acquires temperature
corrections too. They originate from
the `new' radiation \cite{reheat}
which emerges from the decaying
inflaton and has temperature
$T\sim (Hm_PT_{\rm reh}^2)^{1/4}$.
Their scale is set by the thermally
induced mass for the PQ field. A typical
contribution ($\sim\beta NT/m_P$) to
this mass is given by the second term
in the RHS of Eq.~\eq{superpotential}
with the Higgsino lines connected to
a loop via a mass insertion. However,
this contribution exists only under the
condition that the effective Higgsino
mass ($\sim\beta N^2/m_P$) does not
exceed $T$. For our input numbers (see
section~\ref{sect:analysis}) and almost
until the end of the period when
$H\gtrsim m_{3/2}$, we can show that,
whenever this condition is satisfied,
we also have $\beta NT/m_P\lesssim H$,
which means that the temperature
corrections are overshadowed by the
SUGRA ones. At later times, $N$
becomes anyway adequately small and,
thus, the thermally induced mass for
the PQ field comes out much smaller
than $H$. We, thus, conclude that the
temperature corrections can be
neglected throughout the era of
inflaton oscillations.

\par
After reheating, where $H\ll m_{3/2}$
as a consequence of the gravitino
constraint ($T_{\rm reh}
\lesssim 10^9~{\rm GeV}$) \cite{ekn}, the
(one-loop) temperature corrections to the
PQ potential could be the main source of
SUSY breaking. However, as shown in
Refs.~\cite{jean,talks}, these corrections
are less important than the ones from the
hidden sector. This is understood from
the fact that $N$, $\bar{N}$
possess only non-renormalizable
interactions which are suppressed by $m_P$
and carry relatively small dimensionless
coupling constants.

\par
In our case, the SUSY breaking corrections
to the PQ potential during the inflationary
and oscillatory phases can generally be
\cite{drt95} of two types:
\begin{equation}
H^2m_P^2f(\p^2/m_P^2),
\label{mterm}
\end{equation}
\begin{equation}
Hm_P^3g(\p^4/m_P^4),
\label{aterm}
\end{equation}
where $f$, $g$ are some functions depending
on the model, the K\"{a}hler potential and
the phase. To the leading approximation,
the correction in Eq.~(\ref{mterm}) yields
a mass term $\delta m_{\p}^2\p^2/2$ for the
curvaton  $\p$. For a general K\"{a}hler
potential, $\delta m_{\p}^2\sim H^2$ with
either sign possible. However, for specific
(no-scale like) forms of the K\"{a}hler
potential, this induced ${\rm mass}^2$
might be (partially) cancelled
\cite{noscale}. For simplicity, we will
approximate the correction in
Eq.~(\ref{mterm}) by the mass term. We
will also assume that $\delta m_{\p}^2$ is
positive and somewhat suppressed. We,
thus, write
\begin{equation}
\delta m_{\p}=\gamma H,
\label{eq:effec-mass}
\end{equation}
where $\gamma\sim 0.1$ and can have different
values during inflation and field oscillations.
The correction in Eq.~(\ref{aterm}) is of the
$A$ term type with $A\sim H$ and, although it
is generally present, we will not take it into
account in order to simplify the analysis. We
will, thus, consider the following full scalar
potential
\begin{equation}
V(\p)=V_{\rm PQ}+\frac{1}{2}\gamma^2 H^2\p^2+
V_0.
\label{eq:full-pot}
\end{equation}

\par
The evolution of the field $\p$ is governed by
the classical equation of motion
\begin{equation}
\ddot{\p}+3 H \dot{\p}+V'(\p)=0,
\label{eq:field-eqn}
\end{equation}
where overdots and primes denote derivation
with respect to the cosmic time $t$ and the
field $\p$ respectively. Here, the damping
term, proportional to the Hubble parameter,
arises from the expansion of the universe.
For large values of $\p$, the potential
$V_{\rm PQ}$ in Eq.~\eq{PQ-pot} is dominated
by the non-renormalizable $\p^6$-term.
Moreover, for $H\gg\mg$, the induced mass
in Eq.~\eq{effec-mass} can also be important.
Hence, if $\p$ happens to be large during
some stage of inflation, we can write
\begin{equation}
V(\p)\simeq\frac{1}{2}\gamma^2H_{\rm infl}^2
\p^2+\l^2\frac{\p^6}{32 m_P^2}+V_0,
\label{eq:inf-pot}
\end{equation}
where $H_{\rm infl}$ is the inflationary
Hubble parameter. It is natural to assume
that, at the onset of inflation, $V(\p)
\lesssim V_{\rm infl}$, where
$V_{\rm infl}$ is the (almost) constant
vacuum energy density which is provided
by the inflaton and drives the
exponential expansion of the universe.
Thus, $\p$  may initially be
quite large. For such values of $\p$,
$V''$ is dominated by the $\p^6$-term
in the potential in Eq.~\eq{inf-pot}
and the system is initially underdamped
($V''\gg H_{\rm infl}^2$). Consequently,
$\p$ decreases rapidly and $V''$ quickly
falls below $H_{\rm infl}^2$ with the
system entering into a regime of slow-roll
evolution. Soon after this, the induced
mass term in Eq.~\eq{inf-pot} becomes
dominant.

\par
The classical equation of motion holds until
the quantum perturbations of $\p$ generated
by inflation, $\delta\p=H_{\rm infl}/2\pi$,
overshadow its classical kinetic energy
density $\rho_{\rm kin}=\dot{\p}^2/2$. These
perturbations contribute a `quantum' kinetic
energy density of order
$(\delta\p/\delta t)^2/2$, where $\delta t
\sim H_{\rm infl}^{-1}$. Therefore, the
slow-roll regime will end when
$\rho_{\rm kin}\sim H_{\rm infl}^4/8\pi^2$.
From the slow-roll equation $\dot{\p}
\simeq -V'(\p)/3H_{\rm infl}$ derived from
Eq.~\eq{field-eqn}, we then find that this
happens \cite{ls} at $\p\sim\p_Q$, where
$\p_Q$ is given by
\begin{equation}
V'(\p_Q)\simeq\gamma^2H_{\rm infl}^2\p_Q
\sim\frac{3 H_{\rm infl}^3}{2\pi}.
\label{eq:asymptotic}
\end{equation}
The subsequent evolution of $\p$ is
controlled by its quantum perturbations
which `kick' it around to perform a
`random walk'. This results to a `quantum
drift' of the average value of $\p$
towards smaller values (since it cannot
be pushed over $\p_Q$) in the sense that
the field's distribution spreads towards
the origin, which is finally engulfed
inside an evenly spread stationary
distribution whose width is $\p_Q$. Thus,
the value $\p_f$ of $\p$ at the end of
inflation, taken positive without loss
of generality, can be smaller than
$\p_Q$ if sufficiently many e-foldings
take place after $\p_Q$, but it is
typically $\sim \p_Q$. At the termination
of inflation, the quantum perturbations
of $\p$ cease and the field emerges with
vanishing kinetic energy.

\par
Of course, depending on the total number of
e-foldings during inflation and the value
of $\p$ at the onset of inflation, the
inflationary phase may be terminated while
$\p$ is still in the slow-roll regime. In
this case, we have $\p_f\gtrsim\p_Q$ and
$\dot{\p}_f\simeq -\gamma^2H_{\rm infl}
\p_f/3-\l^2\p_f^5/16m_P^2H_{\rm infl}$.
For the values used here, the
corresponding kinetic energy density is
much smaller than the potential energy
density and can be neglected. In summary,
the field $\p$ can practically have any
value $\p_f$ at the end of inflation.
However, if $\p_f$ is smaller than a
certain large limit (depending on $\l$)
so that $\p$ is slowly rolling or in
the `quantum' regime as inflation ends,
the field $\p$ emerges with vanishing
velocity. This sets the initial conditions
for the subsequent evolution of the PQ
field.

\par
After inflation, the universe enters
into a matter dominated era where the
energy density is provided by the
coherent oscillations of the inflaton.
The potential $V(\p)$ is then
given by Eq.~\eq{full-pot} with
$H=2/3t$. If $\p$ is slowly rolling at
the end of inflation, the friction term
in Eq.~\eq{field-eqn} remains important
at subsequent times too and the
evolution of $\p$ is slow for quite some
time. As the cosmic time increases, the
Hubble parameter decreases and, thus, at
some point, the induced mass in
Eq.~\eq{effec-mass} becomes $\lesssim\mg$.
After this point, the soft terms arising
from SUSY breaking by the energy density
of the oscillating inflaton are smaller
than the ones from hidden sector SUSY
breaking. Therefore, the potential can
be approximated as
\begin{equation}
V(\p)\simeq V_{\rm PQ}+V_0,
\label{eq:rad-pot}
\end{equation}
with $V_{\rm PQ}$ given in Eq.~\eq{PQ-pot},
and has a local minimum at the origin with
$V=V_0$, absolute minima at $\p=\pm f_a$
with $V=0$ and local maxima at $\p=\pm
\p_{\rm max}$, where
\begin{equation}
\p^2_{\rm max}=\frac{2}{3\l}\left(|A|-
\sqrt{|A|^2-12}\right)\mg m_P
\label{eq:max}
\end{equation}
with
\begin{equation}
V_{\rm max}=\frac{1}{27\l}
(|A|^2-12)^{\frac{3}{2}}\mg^3m_P
\label{eq:vmax}
\end{equation}
and $\p_{\rm max}>0$. Soon after this,
$H$ becomes $\lesssim\mg$ and the system
enters into a phase of damped oscillations
about either the trivial minimum at $\p=0$
or one of the PQ minima at $\p=\pm f_a$,
depending on the value of $\p$ at
$t\sim\mg^{-1}$.

\par
After reheating, $H=1/2t\ll\mg$.
So, the system remains underdamped and
$V(\p)$ is still given by Eq.~\eq{rad-pot}.
(Recall that the temperature corrections
to the potential are not important.)
The field $\p$ keeps performing damped
oscillations about one of the minima
until it finally decays. These minima are
separated by sizeable potential barriers
and, thus, a transition from the trivial
to a PQ minimum cannot occur. So, if the
field was driven to the trivial minimum,
it would be trapped there forever. The
PQ symmetry would then remain unbroken
and the strong CP problem would not be
solved, while the universe would
experience an unacceptably large
effective cosmological constant.

\section{The PQ field as curvaton}
\label{sect:curvaton}

\par
In accordance with the recently
investigated \cite{curv1,luw02,curv2,
dimo,curv3,gl,curv4,prep} curvaton
scenario, the curvature perturbations
may originate from the quantum
perturbations during inflation not of
the inflaton but rather of another
light field which has been called
curvaton. In this section, we will
address the question whether our PQ
field $\p$ can successfully act as
a curvaton. To this end, we must first
consider the following two requirements:
\begin{enumerate}
\item
As already mentioned, the energy density
of the universe during inflation must
be dominated by $V_{\rm infl}$, the
vacuum energy density provided by the
inflaton. (Here, $V_{\rm infl}$ is
assumed to be essentially constant.)
Therefore, the value of the PQ field
during and at the end of inflation
should be bounded above by the
requirement $V(\p)\lesssim V_{\rm infl}$.
For definiteness, we take here the
standard SUSY hybrid inflationary scheme
\cite{cllsw,shi} (for a recent review see
e.g. Ref.~\cite{talks}). The inflationary
scale is then given by
\begin{equation}
V^{\frac{1}{4}}_{\rm infl}=
\kappa^{\frac{1}{2}}M,
\label{eq:kappa}
\end{equation}
where $\kappa$ and $M$ are the
dimensionless parameter and the mass
scale of the standard superpotential
for hybrid inflation respectively. For
simplicity, we identify $M$ with the
SUSY GUT scale
$M_{\rm GUT}\simeq 2.86\times 10^{16}~
{\rm GeV}$.
\item
A much more stringent constraint is
obtained from the requirement that the
field $\p$ is effectively massless
during (at least) the last $50-60$
inflationary e-foldings so that it
receives a superhorizon spectrum of
perturbations \cite{quantum}
\begin{equation}
\delta \p=\frac{H_{\rm infl}}{2\pi}
\label{eq:fluct}
\end{equation}
from inflation.
This requirement means that $V''(\p)
\lesssim H_{\rm infl}^2$, which
guarantees also that $\p$ is of
negligible time derivative at the
end of inflation.
\end{enumerate}

\par
The detailed evolution of the inflationary
perturbations of $\p$ after inflation will
be considered in the next section. For the
moment, we will only discuss how these
perturbations can generate (or affect) the
total curvature perturbation
\begin{equation}
\zeta=-H\frac{\delta\rho}{\dot\rho}=
\frac{\delta\rho}{3(\rho+p)},
\end{equation}
where $\rho$ and $p$ are the total energy
density and pressure in the universe
respectively. As we approach the time at
which the curvaton decays, the universe
is normally dominated by radiation and the
oscillating curvaton $\p$. In this case,
we can write
\begin{equation}
\zeta=(1-f)\zeta_r+f\zeta_\p,
\end{equation}
where $\zeta_r=\delta\rho_r/4\rho_r$, \
$\zeta_\p=\delta\rho_\p/3\rho_\p$ and
\begin{equation}
f=\frac{3\rho_{\p}}{3\rho_{\p}+4\rho_r}
\label{eq:f}
\end{equation}
with $\rho_r$ and $\rho_\p$ being the
radiation and $\p$ energy densities
respectively. Here, we assumed
that the amplitude of the oscillating $\p$
has been adequately reduced so that the
potential can be approximately considered as
quadratic. The oscillating curvaton then
resembles pressureless matter and $\delta
\rho_\p/\rho_\p=2\,\delta\p_0/\p_0$, where
$\p_0$ is the amplitude of the oscillations
and $\delta\p_0$ the perturbation in this
amplitude originating from $\delta\p$ in
Eq.~\eq{fluct}. According to the curvaton
hypothesis, the inflaton contribution
to the density perturbation is insignificant.
Consequently, the radiation emerging from
the inflaton decay has $\zeta_r\simeq 0$.
After the curvaton decay, we obtain the
final curvature perturbation which, in the
instantaneous decay approximation, is given
by
\begin{equation}
\zeta\simeq f_{\rm dec}\zeta_\p,
\label{eq:fdec}
\end{equation}
where $f_{\rm dec}$ is the value of $f$
at the decay of the curvaton. The COBE
measurements require \cite{cobe} that
$\zeta\simeq 5\times 10^{-5}$.

\par
If the curvaton field decays well before
dominating the energy density of the universe
(i.e. $f_{\rm dec}\ll 1$), which is the case
here, the total curvature perturbation can
acquire \cite{luw02} a significant
non-Gaussian component. The reason is that
the curvature perturbation is proportional
to the perturbation of the curvaton energy
density (see Eq.~\eq{fdec}), which depends
on the square of $\p$. This is interesting
since the recent CMBR data from the WMAP
satellite leave \cite{nongauss} considerable
room for non-Gaussianity in the curvature
perturbations. We will come back to this
point in the next section where a detailed
numerical analysis of the model will be
presented.

\par
As mentioned in section~\ref{sect:evolution},
after the end of inflation, the inflaton
performs damped oscillations and eventually
decays into light particles reheating the
universe. The reheat temperature $T_{\rm reh}$
is related \cite{lectures} to the inflaton
decay width $\Gamma_{\rm infl}$ by
\begin{equation}
T_{\rm reh}=\left(\frac{45}{2\pi^2 g_*}
\right)^{\frac{1}{4}}
(\Gamma_{\rm infl}m_P)^{\frac{1}{2}},
\label{eq:reheat}
\end{equation}
where $g_*$ is the effective number of degrees
of freedom. After reheating, gravitinos
($\tilde{g}$) are thermally produced besides
other particles. They only decay, though, well
after the big bang nucleosynthesis (BBN) since
they have very weak couplings. The decay
channels normally include the one to a photon
and a photino ($\tilde{g}\rightarrow\gamma+
\tilde\gamma$) with the emerging high energy
photons causing the dissociation of the light
elements and, thus, destroying the successful
predictions of BBN. This yields \cite{ekn}
the strongest possible upper bound on the
reheat temperature, which is $T_{\rm reh}
\lesssim 10^9~{\rm GeV}$ if the branching
ratio to photons equals unity and the
gravitino mass $m_{3/2}$ is in the range of
several hundreds of GeV. However, the upper
bound can be considerably relaxed
\cite{ekn,km} if this branching ratio is
less than unity.

\par
Each decaying gravitino yields one sparticle
subsequently turning into the lightest
sparticle (LSP), which is stable. This is
due to the presence of `matter parity' in
our model. Moreover, no pair annihilation of
the resulting LSPs is possible since the
universe is already quite cold. Consequently,
these LSPs survive until the present time
contributing to the relic abundance of CDM
in the universe. We will assume, as it is
customary, that the LSP is the lightest
neutralino ($\tilde\chi$). We will also make
the simplifying assumption that the relic
abundance of the thermally produced LSPs is
negligible, which holds in many cases. So,
the relic abundance, $\Omega_{\rm LSP}h^2$,
of the LSPs comes solely from the decaying
gravitinos and can be estimated following
Ref.~\cite{km}. Here, $\Omega_{\rm LSP}=
\rho_{\rm LSP}/\rho_c$ with $\rho_{\rm LSP}$
($\rho_c$) being the LSP (critical) energy
density and $h\simeq 0.71$ is the present
Hubble parameter in units of $100~\rm{km}~
\rm{sec}^{-1}~\rm{Mpc}^{-1}$.

\par
Our model contains axions which can also
contribute to the CDM of the universe.
Their relic abundance, $\Omega_{a}h^2$, can
be calculated by applying the formulae of
Ref.~\cite{turner} ($\Omega_a=\rho_a/\rho_c$
with $\rho_a$ being the axion energy density).
For simplicity, we take
the QCD scale $\Lambda=200~{\rm GeV}$ and
ignore the uncertainties in these formulae.
Also, the anharmonic effects are negligible
in our case and the possibility of entropy
production after the QCD transition is not
considered. The total CDM abundance is the
sum of the relic abundance of the LSPs from
the gravitino decays and the relic axion
abundance ($\Omega_{\rm CDM}=\rho_{\rm CDM}/
\rho_c$ with $\rho_{\rm CDM}$ being the CDM
energy density):
\begin{equation}
\Omega_{\rm CDM}h^2\simeq 0.0074\left(\frac
{m_{\rm LSP}}{200~{\rm GeV}}\right)\left(
\frac{T_{\rm reh}}{10^{9}~{\rm GeV}}\right)+
\theta^2\left(\frac{f_a}{10^{12}~{\rm GeV}}
\right)^{1.175},
\label{eq:oh2}
\end{equation}
where $m_{\rm LSP}$ is the LSP mass and
$\theta$ the so-called initial misalignment
angle, i.e. the phase of the complex PQ
field at the end of inflation. The angle
$\theta$ lies in the interval
$[-\pi/{\cal N},\pi/{\cal N}]$,
where ${\cal N}$ is the absolute value of
the sum of the PQ charges of all fermionic
color (anti)triplets. This same ${\cal N}$
determines the $Z_{\cal N}$ subgroup of
$U(1)_{\rm PQ}$ which remains unbroken by
instantons. As mentioned
in section~\ref{sect:solution}, in our (and
the simplest) model, ${\cal N}=6$. All
$\theta$'s in the above interval are equally
probable.

\par
It is well-known that all the effectively
massless (non conformally invariant)
fields acquire a superhorizon spectrum of
perturbations during inflation. So, such
perturbations are
generated in the axion field too. When the
axion mass comes into play at the QCD phase
transition, these perturbations become
isocurvature axion density perturbations.
The late universe consists, in our case, of
axions ($a$), baryons ($B$) and neutralinos
($\tilde\chi$) as matter, and photons
($\gamma$) and (massless) neutrinos ($\nu$)
as radiation. In order to extract the
restrictions on our model which come from
isocurvature perturbations, we employ the
analysis of Ref.~\cite{agws}. We, thus,
consider the comoving curvature perturbation
$\mathcal{R}_{\rm rad}$ and the isocurvature
perturbation $\mathcal{S}_{\rm rad}$
determining the large scale CMBR temperature
perturbation. They are given by
\begin{eqnarray}
\mathcal{R}_{\rm rad}&=&-\zeta,\\
\label{R}
\mathcal{S}_{\rm rad}\hat{a}&=&
\sum_{i=a,B,\tilde\chi}
\frac{\Omega_i}{\Omega_m}S_i\hat{a}_i,
\label{S}
\end{eqnarray}
where $S_i=\delta n_i/n_i-3\,\delta T/T$
($i=a,~B,~\tilde\chi$) is the isocurvature
perturbation in the $i$th species, $\hat{a}$,
$\hat{a}_i$ are appropriate Gaussian random
variables, $\Omega_i=\rho_i/\rho_c$ and
$\Omega_m=\rho_m/\rho_c$ with $\rho_i$ and
$n_i$ being the energy and number densities
of the $i$th species and
$\rho_{m}=\rho_{a}+\rho_{B}+
\rho_{\tilde\chi}$ being the matter energy
density.

\par
The neutralinos appear as decay products of the
gravitinos which were generated at reheating.
They, thus, inherit the same curvature
perturbation, which coincides with that of
radiation emerging from the inflaton decay.
However, $\zeta_r\simeq 0$ by the curvaton
hypothesis and, consequently,
$\zeta_{\tilde\chi}=\delta\rho_{\tilde\chi}/3
\rho_{\tilde\chi}$ also vanishes. After the
curvaton decay, the neutralino isocurvature
perturbation $S_{\tilde\chi}\equiv 3
(\zeta_{\tilde\chi}-\zeta_\gamma)\simeq -3
\zeta$. Here, we assumed that the neutrino
isocurvature perturbation is negligible and,
thus, $\zeta_\gamma=\delta\rho_\gamma/4
\rho_\gamma\simeq\zeta$ ($\rho_\gamma$ is
the photon energy density). Baryons can be
produced via a primordial leptogenesis
\cite{lepto} which can occur \cite{leptoinf}
at reheating. The inflaton decays into right
handed neutrino superfields which,
subsequently, decay into Higgs and lepton
superfields generating a net lepton number.
As a consequence, $\zeta_B\simeq 0$ too and
again $S_B\equiv 3(\zeta_B-\zeta_\gamma)
\simeq -3\zeta$. We see that the baryons and
the neutralinos carry an isocurvature
perturbation which is fully correlated with
the isentropic perturbation in the notation
of Ref.~\cite{agws}.

\par
The axions are produced at the QCD phase
transition which takes place comfortably
after the curvaton decay. At this transition,
the axion mass $m_a$ is generated and the
axion field starts oscillating. For the
amplitudes $a$ encountered here, the anharmonic
effects \cite{turner} are negligible and, thus,
the axion energy density is given by $\rho_a
=m_a^2a^2/2$. The isocurvature perturbation
which is generated is then $S_a=\delta n_a/n_a
=\delta\rho_a/\rho_a=2\,\delta a/a$. The
perturbation $\delta a$ in the amplitude of the
axion field comes from inflation and can be
estimated \cite{lyth} as follows. Suppose that
$\p_*$ is the value of $\p$ when our present
horizon scale crosses outside the inflationary
horizon (i.e. $\sim 50-60$ e-foldings before
the end of inflation) and $\theta$ is the
initial misalignment angle, i.e. the phase of
the complex PQ field during (and at the end of)
inflation. This angle corresponds to the
canonically normalized real scalar field
$\theta\p_*$ which acquires a perturbation
$H_{\rm infl}/2\pi$ from inflation and, thus,
$\delta\theta=H_{\rm infl}/2\pi\p_*$. The
initial amplitude of axion oscillations $a=
\theta f_a$ then acquires a perturbation
$\delta a=H_{\rm infl}f_a/2\pi\p_*$. We see
that the axion isocurvature perturbation is
$S_a=H_{\rm infl}/\pi\theta\p_*$. This
perturbation is completely uncorrelated
with the curvature perturbation.

\par
Replacing $S_a$, $S_B$ and $S_{\tilde\chi}$ in
Eq.~(\ref{S}), we obtain
\begin{equation}
\mathcal{S}_{\rm rad}\hat{a}=
\left(\frac{H_{\rm infl}}{\pi\theta\p_*}
\frac{\Omega_a}{\Omega_m}\hat{a}_a-
3\zeta\frac{\Omega_B+\Omega_{\tilde\chi}}
{\Omega_m}\hat{a}_B\right),
\end{equation}
which consists of an uncorrelated and a fully
correlated part. Here, the Gaussian random
variables $\hat{a}_a$, $\hat{a}_B=
\hat{a}_{\tilde\chi}$ are independent. The
dimensionless cross correlation parameter is
given by \cite{agws}
\begin{equation}
\cos\Delta=\frac{3}{B}\left(1-\frac{\Omega_a h^2}
{\Omega_m h^2}\right),
\label{eq:cross-corr}
\end{equation}
where
\begin{equation}
B=\left[\left(\frac{H_{\rm infl}}
{\pi\theta\p_*\zeta}\right)^2\left(
\frac{\Omega_a h^2}{\Omega_m h^2}\right)^2
+9\left(1-\frac{\Omega_a h^2}{\Omega_m h^2}
\right)^2\right]^\frac{1}{2}.
\label{eq:isoc-bound}
\end{equation}
is the entropy-to-adiabatic ratio. In
Ref.~\cite{agws}, a bound on this ratio has
been derived from the pre-WMAP CMBR data
and was depicted in Fig.~2 of
this reference versus $\cos\Delta$ with the
other parameters marginalized. We take an
improved version \cite{private} of this
figure, which can be derived from the
extended set of pre-WMAP data used in
Ref.~\cite{gl}, and extract from it the
$95\%$ confidence level (c.l.) bound on $B$
for each value of $\cos\Delta$. At present,
this is the most stringent available bound
on isocurvature perturbations of mixed
correlation with the curvature perturbation.

\section{Numerical analysis and results}
\label{sect:analysis}

\subsection{The band structure}
\label{subsect:band}

\par
We are now ready to study numerically the
evolution of the PQ field after inflation.
To this end, we perform the change of
variable
\begin{equation}
t=H_{\rm infl}^{-1} e^{\tau}
\end{equation}
in Eq.~\eq{field-eqn}, which then becomes
\begin{equation}
\frac{d^2\p}{d\tau^2}+\left(3e^{\tau}
\frac{H}{H_{\rm infl}}-1\right)
\frac{d\p}{d\tau}+\frac{e^{2\tau}}
{H_{\rm infl}^2} V'(\p)=0.
\label{eq:num-eqn}
\end{equation}
Inflation ends at $t=t_f\simeq (2/3)
H_{\rm infl}^{-1}$. So, we impose
the initial conditions $\p=\p_f>0$
and $d\p/d\tau=0$ at $\tau=\ln (2/3)$.
Although, as explained in
section~\ref{sect:evolution},
the field $\p$ exits inflation with
negligible velocity only if $\p_f$ is not
too large, we will apply these initial
conditions uniformly for all values of
$\p_f$ in order to obtain an overall
picture of the possible behavior patterns
of the system. We will, of course, use
only the relevant $\p_f$'s to draw
physical conclusions. As we already
mentioned,
these $\p_f$'s practically coincide
with the $\p_f$'s for which the PQ field
is effectively massless during inflation
so that it acquires a superhorizon
spectrum of perturbations.

\par
The Hubble parameter during inflation
$H_{\rm infl}$ is given by the Friedmann
equation
\begin{equation}
H_{\rm infl}^2=\frac{V_{\rm infl}}
{3m_P^2}=\frac{\kappa^2 M_{\rm GUT}^4}
{3m_P^2},
\label{eq:inf-hubble}
\end{equation}
where the almost constant energy density
$V_{\rm infl}$ driving inflation is taken
from Eq.~\eq{kappa} which holds in the
case of standard SUSY hybrid inflation.
We do not commit ourselves to any details
of this inflationary scheme. We also
leave out the possible restrictions from
the specific reheating and leptogenesis
processes in the SUSY GUT models of
Ref.~\cite{inf}. We just use, for our
analysis, the inflationary scale derived
from Eq.~\eq{kappa} as a working example.
For definiteness, we
further take $\kappa=10^{-3}$, which
gives $V_{\rm infl}^{1/4}\simeq 9.04
\times 10^{14}~{\rm GeV}$ and
$H_{\rm infl}\simeq 1.97\times 10^{11}~
{\rm GeV}$. It should be noted that this
inflationary scale is consistent with
the curvaton hypothesis which requires
\cite{dimo} that $V^{1/4}_{\rm infl}
\lesssim 3\times 10^{15}~{\rm GeV}$.
Indeed, if this condition holds, the
curvature perturbation from the inflaton
can be made negligible.

\par
The scalar potential $V(\p)$ is taken
from Eq.~\eq{full-pot}, where we put
$\mg=300~{\rm GeV}$, $|A|=5$ and
$\gamma=0.1$. The parameter $\lambda$ is
taken to range between $10^{-4}$ and
$10^{-3}$. Fixing the reheat temperature
$T_{\rm reh}$ to the value
$10^9~{\rm GeV}$
\cite{ekn} and taking $g_*=228.75$ which
corresponds to the MSSM spectrum,
Eq.~\eq{reheat} yields the inflaton decay time
$t_{\rm reh}=
\Gamma_{\rm infl}^{-1}\simeq 2.44\times
10^{-1}~{\rm GeV}^{-1}$. For $t_f\leq t
\leq t_{\rm reh}$,
the universe is dominated by the
oscillating inflaton and $H=2/3t$. The
expression in parentheses in
Eq.~\eq{num-eqn} is then equal to unity.
For $t\geq t_{\rm reh}$, however,
radiation dominates and
$H=1/2(t-t_{\rm reh}/4)$. Here, we
subtracted $t_{\rm reh}/4$ from $t$
in order to achieve continuity at
$t=t_{\rm reh}$. For $t\gg t_{\rm reh}$,
the expression in parentheses in
Eq.~\eq{num-eqn} equals $1/2$.

\FIGURE{\epsfig{file=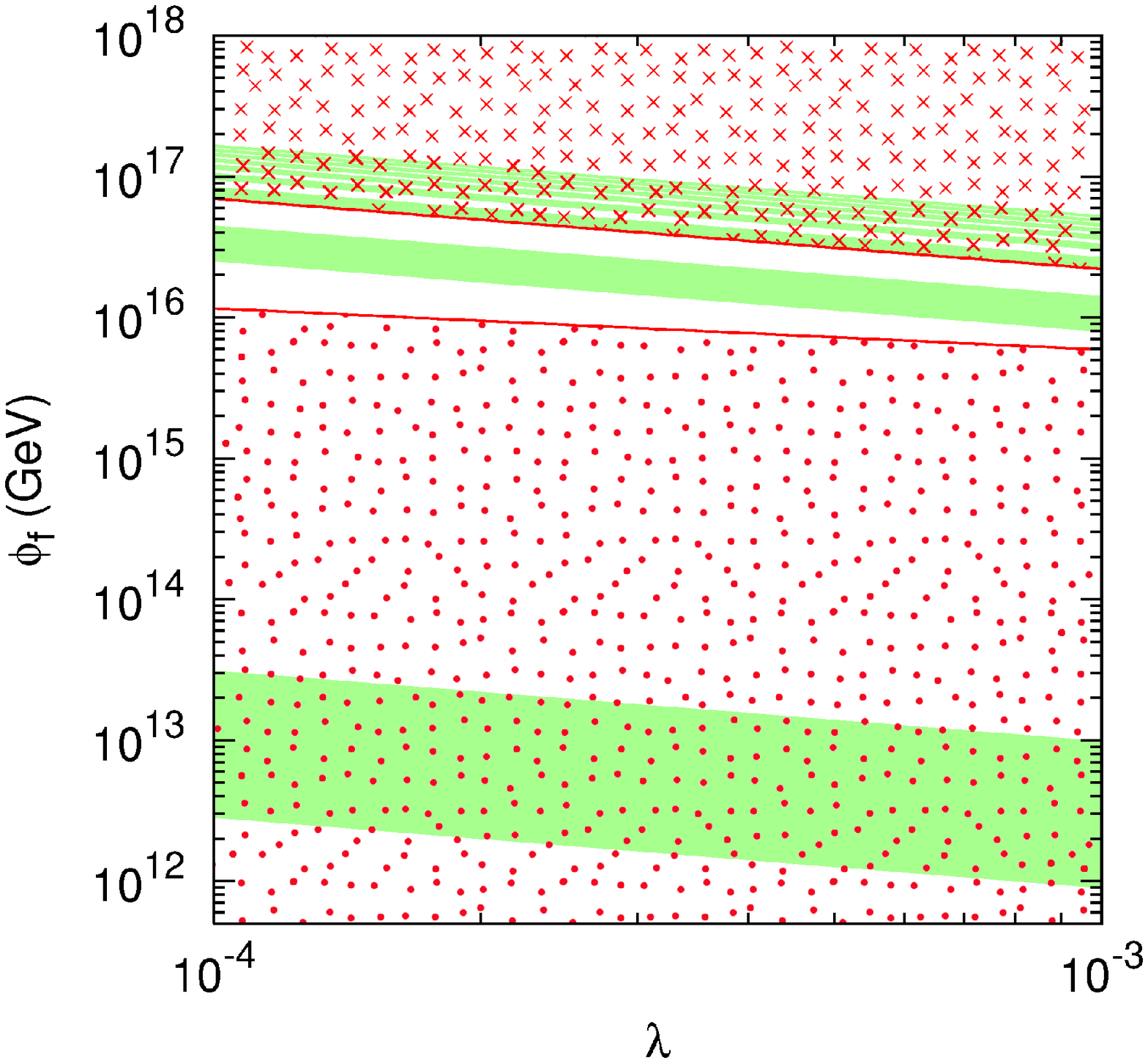,
width=12.04cm}
\label{fig:bands}
\caption{The band structure in the
$\l-\p_f$ plane. If $\p_f$ lies in a green
(white) band, the system is led into a PQ
(the false) vacuum. The red crossed area is
excluded by requiring that the PQ field is
light
during the relevant part of inflation, while
the red dotted area is excluded by the bound
on the isocurvature perturbation from the
CMBR data.}
\label{bands}}

\par
Using the above described initial
conditions and numerical inputs, we now
solve Eq.~\eq{num-eqn} numerically. For
each value of $\p_f$ and any given $\l$
in the range $10^{-4}-10^{-3}$, we follow
the evolution of the system and identify
the minimum of $V_{\rm PQ}$ about which
the PQ field is finally set in coherent
damped oscillations. We find that a
well-defined band structure emerges in
the $\l-\p_f$ plane. This structure is
depicted in Fig.~\ref{bands} with
alternating white and green bands. We
number the successive bands starting
from zero which is given to the lowest
lying (white) band. So, the white bands
correspond to even integers, while the
green ones to odd integers. We only show
the first 18 bands since after this the
bands become too dense in the figure to
be clearly distinguishable. If
the initial value of the field $\p_f$
belongs to a white band, the system is
finally led into the false vacuum at
$\p=0$. So, such $\p_f$'s are excluded.
However, if $\p_f$ lies in a green band,
the system ends up in one of the PQ
vacua.  The boundaries of the bands in
Fig.~\ref{bands}, which is a $\log-\log$
plot, are parallel straight lines with
an inclination equal to $-1/2$. This is
easily understood by observing that, if
$\p$ is rescaled by absorbing a
$\l^{1/2}$ factor, Eq.~\eq{num-eqn}
becomes $\l$-independent.

\par
As indicated in section~\ref{sect:curvaton},
the PQ field must be light during the
last $50-60$ e-foldings (i.e. $V''(\p)
\lesssim H_{\rm infl}^2$) so that it
acquires perturbations from inflation which
enable it to act as curvaton. In particular,
at the end of inflation, we must require
that $V''(\p_f)\lesssim H_{\rm infl}^2$,
which excludes the red crossed area in
Fig.~\ref{bands}. The boundary of this area
is again a straight line with an inclination
equal to $-1/2$ and turns out to be very
close to the lower boundary of the 5th
(green) band. So, out of all the odd (green)
bands, only the 1rst and 3rd are consistent
with the above requirement, while all
the higher green bands are excluded. We
find that, in the non-excluded area, $\p$
remains practically constant in the last
$50-60$ e-foldings and, thus, the condition
$V''(\p)\lesssim H_{\rm infl}^2$ is valid
throughout this period. It is obvious that,
in this same area, the field velocity at
the end of inflation is negligible. Finally,
we should note that, in this allowed area,
the condition $V(\p)\lesssim V_{\rm infl}$,
which, as explained in
section~\ref{sect:curvaton}, must hold
during inflation, is automatically
satisfied provided that the total number
of inflationary e-foldings is not huge.

\par
It should be emphasized that the
non-vanishing of $N$, $\bar{N}$ after
inflation alone cannot guarantee the
avoidance of disastrous domain walls.
It is crucial to further ensure that
the system finally relaxes in the
same minimum over all space. This
is readily fulfilled in our
case since the perturbation $\delta\p=
H_{\rm infl}/2\pi\simeq 3.14\times
10^{10}~{\rm GeV}$, which is acquired
by the PQ field during inflation, is
much smaller than the width of the
bands. Therefore, right after
inflation, the field is found to lie
within the same band everywhere in
space. It is then led into the same
minimum over all space and no
catastrophic domain wall production
occurs. The opposite situation has
been encountered in Ref.~\cite{ls96},
where the inflationary perturbation of
the field could become comparable to
or even larger than the width of the
bands. As a consequence, topological
defects could be generated.

\par
We will now consider the bound on the
entropy-to-adiabatic ratio $B$ in
Eq.~\eq{isoc-bound} which has been
derived in Ref.~\cite{agws} from the
pre-WMAP CMBR data.
We take the baryon abundance $\Omega_B
h^2=0.02$, which is \cite{bau} its
central value from BBN. The CDM
abundance, $\Omega_{\rm CDM}h^2$, is
taken equal to its central value
$0.12$ from DASI \cite{cdm}. Thus, the
total matter abundance is $\Omega_mh^2
=0.14$. The  axion
abundance $\Omega_ah^2$ (second term in
the RHS of Eq.~\eq{oh2}) can then be
calculated from Eq.~\eq{oh2}, where we
put $m_{\rm LSP}=200~{\rm GeV}$. We find
that the axions constitute about $94\%$
of the CDM. From this, we find the initial
misalignment angle $\theta$ for each
$\l$. The curvature perturbation $\zeta$
is taken equal to $5\times 10^{-5}$ from
COBE \cite{cobe}. As we saw, $\p$ remains
practically constant in the last
$50-60$ e-foldings provided that we
are outside the red crossed area of
Fig.~\ref{bands}. Thus, in the relevant
area, the value $\p_*$ of $\p$ when our
present horizon crossed outside the
inflationary horizon can be identified
with $\p_f$. We apply an improved
version \cite{private} of the $95\%$ c.l.
bound on $B$ for each value of the cross
correlation parameter $\cos\Delta$ in
Eq.~\eq{cross-corr}. This yields a lower
bound on $\p_f$ for each $\l$ excluding
the red dotted area in Fig.~\ref{bands}.
The boundary of this area is a straight
line with an inclination equal to
$-1.175/4\simeq -0.294$ as one can
easily deduce from Eqs.~\eq{fa}, \eq{oh2},
\eq{cross-corr} and \eq{isoc-bound}.

\par
We see that the bound from the
isocurvature perturbation excludes the
1rst (green) band and, thus, only the
3rd (green) band survives. Note,
though, that this bound actually pushes
$\p_{*}$ and not $\p_f$ to large values.
Moreover, in the last $50-60$ e-foldings,
such a large $\p_{*}$ could be reduced
to a $\p_f$ lying in the 1rst band if
$\p$ was rolling fast. However, this
would require a large SUGRA induced
curvaton mass during inflation, which
is incompatible with the curvaton
hypothesis. Finally, we should stress
that the fact that the CDM consists
mainly of axions carrying an
isocurvature perturbation which is
uncorrelated with the curvature
perturbation is of crucial importance
for the viability of the model. Indeed,
it leads to a cross correlation
parameter $\cos\Delta$ which is
somewhat smaller than unity and, thus,
the upper bound on the
entropy-to-adiabatic ratio $B$ is
considerably relaxed. Actually,
in our case, this bound takes its
maximal value ($\simeq 0.95$) which is
achieved in the range $0.55\lesssim
\cos\Delta\lesssim 0.75$.

\subsection{The evolution patterns of
the PQ field}
\label{subsect:pattern}

\par
Let us now describe in some detail the
time dependence of the full curvaton
potential $V(\p)$ in Eq.~\eq{full-pot}
and the evolution of the PQ field $\p$
after the end of inflation in
accordance with our numerical findings.
For definiteness, we fix $\lambda=
10^{-4}$. To obtain the picture for
other values of $\lambda$, we just have
to rescale $V$ as $\l^{-1}$ and $\p$
as $\l^{-1/2}$. In Fig.~\ref{pot-first},
we present a three dimensional plot of
$V(\p)$ as a function of $\p$ and
the cosmic time $t$. We see that,
initially and for small $\p$'s, the
induced mass in Eq.~\eq{effec-mass}
dominates over the soft SUSY breaking
terms in Eq.~\eq{vpq} and, thus,
$V(\p)$ has just one minimum at zero.
At later times, the induced mass
becomes comparable to the soft
terms and $V(\p)$ develops a pair of
symmetric local minima separated from
the trivial minimum (at zero) by a
pair of symmetric local maxima. The
height of the local minima is gradually
reduced and, at some point, they become
absolute minima. Finally, when $t\gtrsim
2\gamma/3m_{3/2}\simeq 2.22\times
10^{-4}~{\rm GeV}^{-1}$, the induced
mass becomes subdominant and, thus,
$V(\p)$ can be approximated as in
Eq.~\eq{rad-pot}. The above
(non-trivial) minima then approach the
PQ vacua at $\p=\pm f_a\simeq\pm 6.48
\times 10^{12}~{\rm GeV}$ with $V=0$.
The height of the local maxima also
decreases as the universe evolves and,
for $t\gtrsim 2\gamma/3m_{3/2}$, they
approach the maxima of the (approximate)
potential in Eq.~\eq{rad-pot}, which lie
at $\p=\pm\p_{\rm max}\simeq\pm 2.61
\times 10^{12}~{\rm GeV}$ with potential
energy density $V=V_{\rm max}\simeq
(3.27\times 10^7~{\rm GeV})^4$ (see
Eqs.~\eq{max} and \eq{vmax}). The
trivial minimum (at $\p=0$) exists at
all times and possesses a constant
potential energy density $V=V_0\simeq
(3.16\times 10^7~{\rm GeV})^4$ (see
Eq.~\eq{v0}).

\FIGURE{\epsfig{file=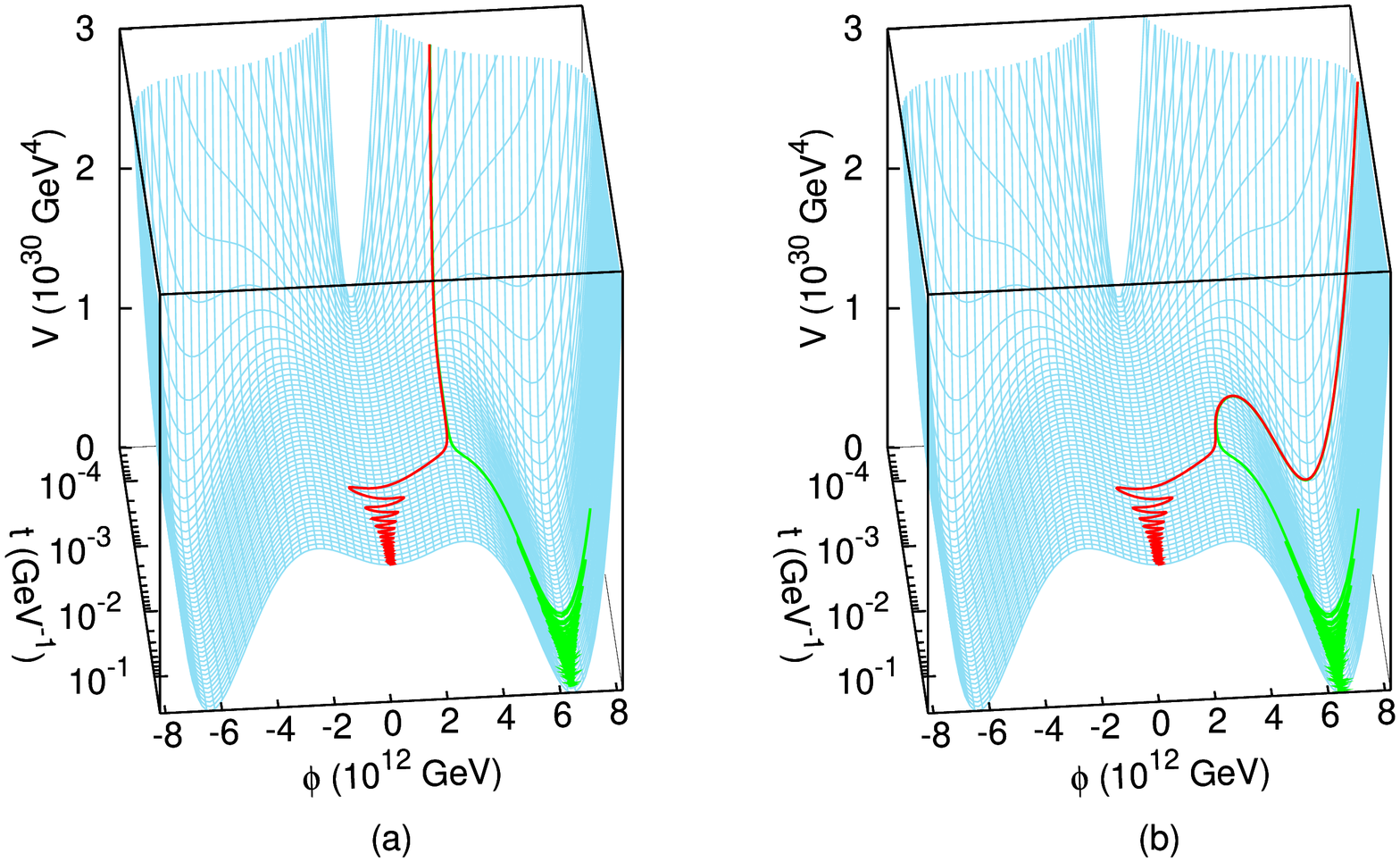,
width=8.426cm,angle=-90}
\caption{The three dimensional plot of
the full curvaton potential $V(\p)$
versus $\p$ and $t$ for $\l=10^{-4}$.
The green (red) line in (a) depicts
the path followed by $\p$ if $\p_f$ is
a little above (below) the lower
boundary of the 1rst band, while the
green (red) line in (b) corresponds to
the field evolution if $\p_f$ is a
little below (above) the upper boundary
of the same band.}
\label{pot-first}}

\par
In order to get a feeling of the
post-inflationary evolution patterns of
$\p$ in the various bands, let us first
consider that $\p_f$ is slightly above
(below) the lower boundary of the 1rst
band which lies at $\p_f\simeq 2.83
\times 10^{12}~{\rm GeV}$ for
$\l=10^{-4}$. The interesting part of
the path followed by $\p$ is depicted
in Fig.~\ref{pot-first}a by a green
(red) line. The field is initially
overdamped and decreases monotonically
at a very slow pace. For
$t\gtrsim 2\gamma/3m_{3/2}\simeq 2.22
\times 10^{-4}~{\rm GeV}^{-1}$, $\p$ is
found to lie slightly above (below)
$\p_{\rm max}\simeq 2.61
\times 10^{12}~{\rm GeV}$. For $t
\gtrsim 2/3m_{3/2}\simeq 2.22\times
10^{-3}~{\rm GeV}^{-1}$, $H\lesssim
m_{3/2}$ and the system becomes
underdamped. It falls towards the PQ
vacuum at $\p=f_a$ (the trivial false
vacuum) and starts performing damped
oscillations about it. Soon after this,
the universe enters into the radiation
dominated era which follows reheating
and the damped oscillations of $\p$
continue until this field finally
decays. For $\p_f\lesssim 2.83\times
10^{12}~{\rm GeV}$, which corresponds
to the zeroth (white) band, the field
always rolls monotonically towards the
false vacuum at zero and, eventually,
is set in damped oscillations about it.

\FIGURE{\epsfig{file=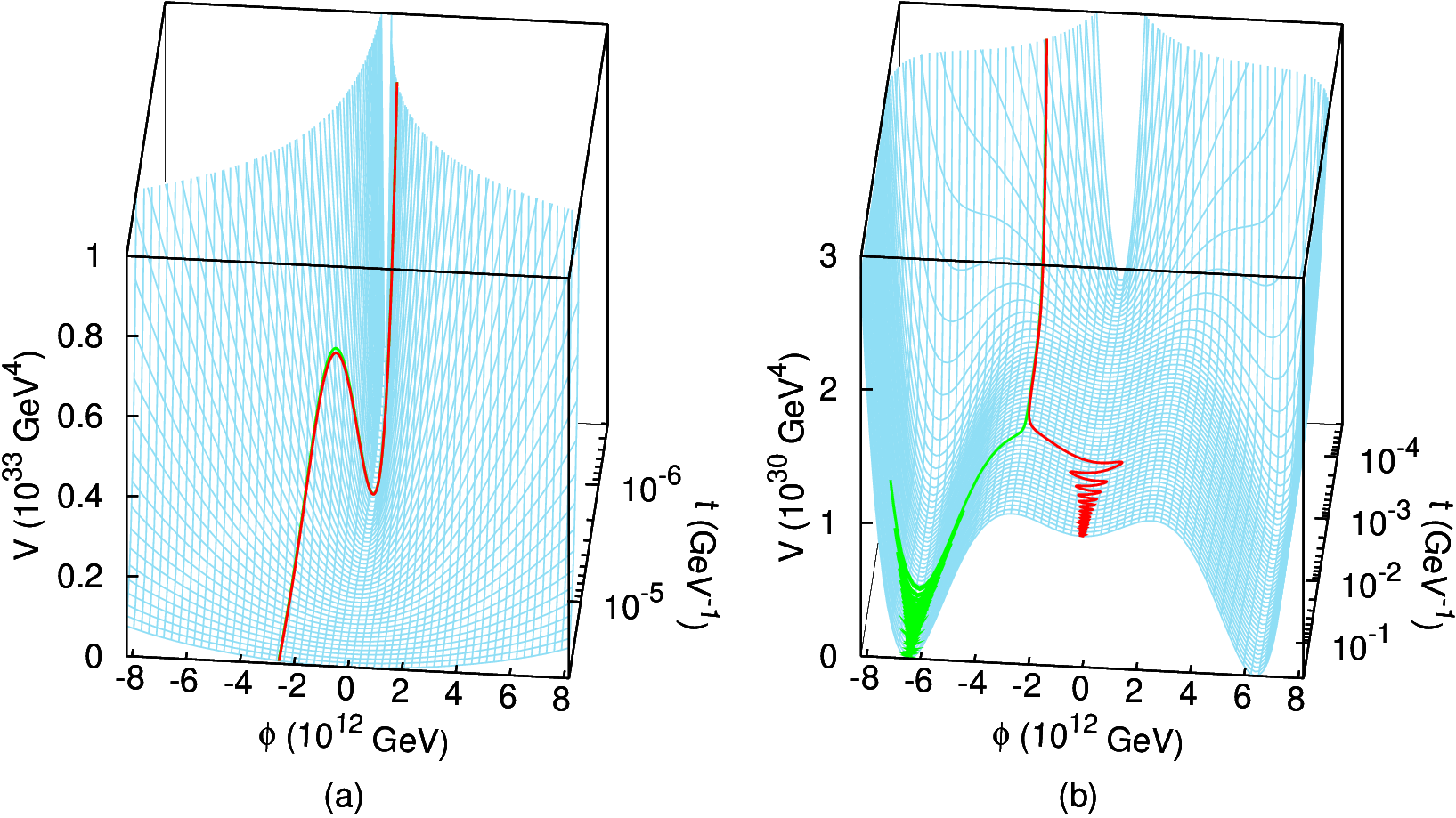,
width=8.426cm,angle=-90}
\caption{The PQ field evolution for
$\p_f$ near the lower boundary of the
3rd (green) band and with $\l=10^{-4}$.
The green (red) line corresponds to
$\p_f$ being slightly above (below)
this boundary. The earlier evolution
is shown in (a), while the later one
in (b).}
\label{pot-thirdI}}

\par
In Fig.~\ref{pot-first}b, we show the
red (green) path followed by the field
$\p$, if $\p_f$ is taken to lie
slightly above (below) the upper
boundary of the 1rst band at $\p_f
\simeq 3.11\times 10^{13}~{\rm GeV}$
for $\l=10^{-4}$. We see that the
field, as it decreases slowly, passes
from the PQ vacuum at $\p=f_a$ and
then climbs up the potential until it
reaches a value slightly below (above)
$\p_{\rm max}$. It
stays there for some time and, finally,
falls towards the trivial vacuum (the
PQ vacuum at $\p=f_a$) and performs
damped oscillations about it. For
$2.83\times 10^{12}~{\rm GeV}\lesssim
\p_f\lesssim 3.11\times 10^{13}~
{\rm GeV}$, which corresponds to the
1rst (green) band, the field always
rolls slowly directly into the PQ
vacuum at $\p=f_a$ and oscillates
about it. We should note that the
upper limit of the `quantum' regime
$\p_Q$, which turns out to be $\simeq
9.41\times 10^{12}~{\rm GeV}$ (see
Eq.~\eq{asymptotic}), lies is this
band for the range of $\l$ considered
here.

\FIGURE{\epsfig{file=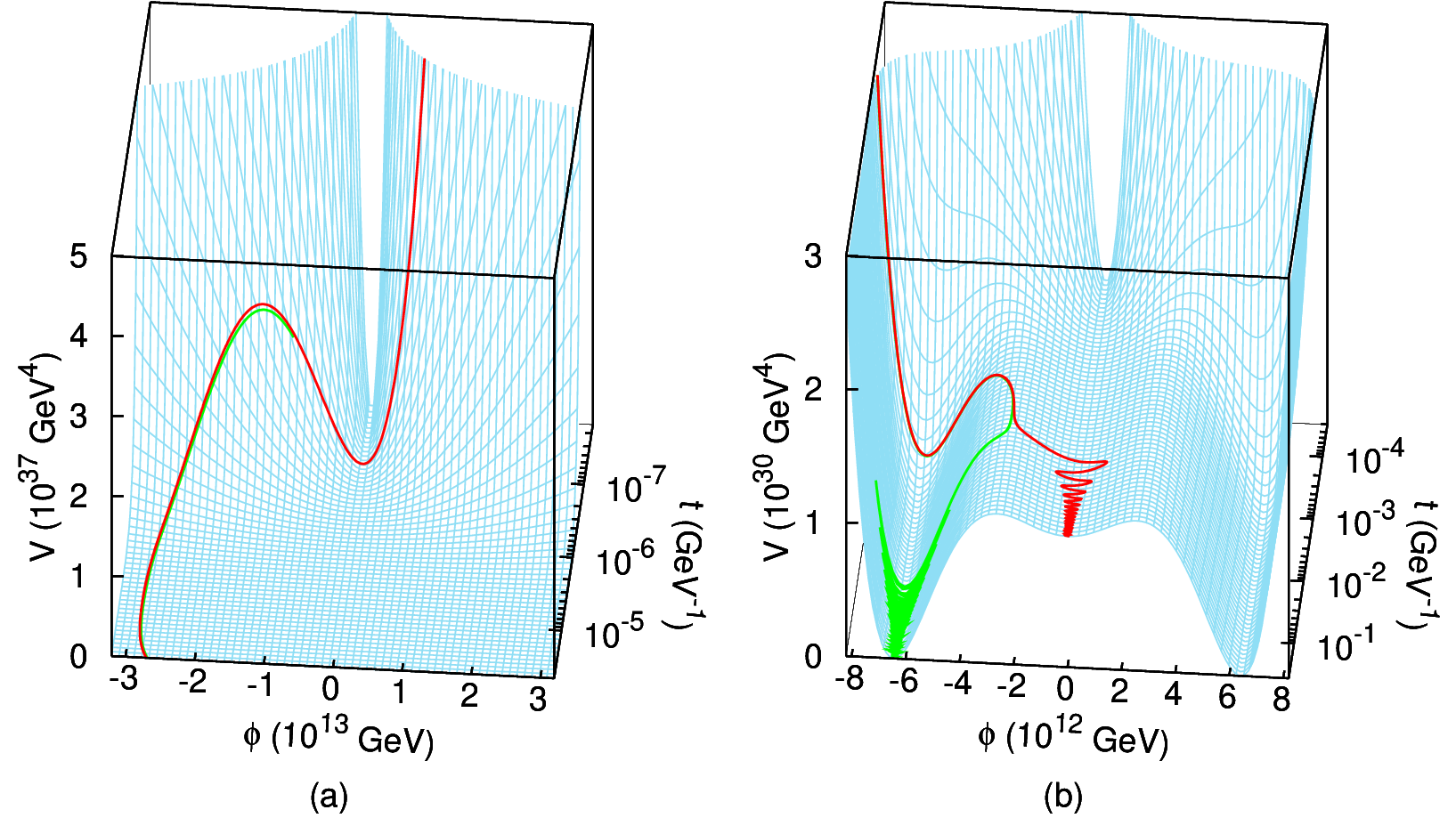,
width=8.426cm,angle=-90}
\caption{The PQ field evolution for
$\p_f$ near the upper boundary of the
3rd (green) band and with $\l=10^{-4}$.
The green (red) line corresponds to
$\p_f$ being slightly below (above)
this boundary. The earlier evolution
is shown in (a), while the later one
in (b).}
\label{pot-thirdII}}

\par
In Fig.~\ref{pot-thirdI}, we present
the evolution of the field $\p$ when
$\p_f$ happens to lie near the lower
boundary of the 3rd (green) band at
$\p_f\simeq 2.5\times 10^{16}~
{\rm GeV}$ again for $\l=10^{-4}$.
The green (red) line corresponds to
$\p_f$ being slightly above (below)
this boundary. The earlier evolution
of the PQ field is depicted in
Fig.~\ref{pot-thirdI}a. We see that,
well before the appearance of the
non-trivial minima, the field rolls
down slowly, reaches the trivial
vacuum and then passes to negative
values slowly climbing up the
potential. It is, eventually,
stabilized for some time slightly
after (before) $-\p_{\rm max}$ and
finally falls
into the PQ vacuum at $\p=-f_a$
(trivial vacuum) as is obvious from
Fig.~\ref{pot-thirdI}b, where the
subsequent evolution is shown. We
find that, for $3.11\times 10^{13}
~{\rm GeV}\lesssim\p_f\lesssim 2.5
\times 10^{16}~{\rm GeV}$, which
corresponds to the 2nd (white) band,
the field always falls into the
trivial vacuum and oscillates about
it.

\par
Fig.~\ref{pot-thirdII} shows the
green (red) trajectory followed by
$\p$ for $\p_f$ slightly below
(above) the upper boundary of the
3rd band at $\p_f\simeq 4.48\times
10^{16}~{\rm GeV}$ for $\l=10^{-4}$.
We see that the slowly moving field
passes from zero, overtakes
$-f_a$, falls back to the PQ
minimum at $\p=-f_a$, which has
appeared by that time, climbs up to
a point slightly below (above)
$-\p_{\rm max}$ and
stays there for a period of time.
It then falls towards the PQ minimum
at $\p=-f_a$ (the trivial minimum)
and is set in damped oscillations.
Our numerical results show that, for
$2.5\times 10^{16}~{\rm GeV}\lesssim
\p_f\lesssim 4.48\times 10^{16}~
{\rm GeV}$, which corresponds to the
3rd (green) band, the field always
falls into the PQ vacuum at $\p=-f_a$
and oscillates about it. More
complicated patterns of evolution of
the PQ field are encountered in the
higher bands, but we will not discuss
them in any detail since these bands
are both physically meaningless and
useless for our purposes as already
explained.

\par
It is interesting to note that the
2nd (white) band extends over almost
three orders of magnitude in $\p_f$
and, thus, pushes the 3rd (green)
band to large values of $\p_f$. Also,
the lower bound on $\p_f$ from the
isocurvature perturbation is quite
sizeable. The combined effect of
these two facts is that the $\p_f$'s
which can be useful for our purposes
are fairly large. As a consequence,
the relative perturbation,
$\delta\p/\p$, acquired by the
curvaton during inflation is
somewhat small. This leads to a
small partial curvature perturbation,
$\zeta_\p$, carried by the
oscillating PQ field at the time of
its decay. So, as can be seen from
Eq.~\eq{fdec}, a relatively large
$f_{\rm dec}$ is required in order
to achieve the COBE constraint on the
total curvature perturbation, which
makes our task a little tougher. The
situation would have certainly been
less tight if the 1rst (green) band
could be used.

\par
One way to understand the fact that
the 2nd band comes out so wide is
the following. Consider the
classical equation of motion in
Eq.~\eq{field-eqn} during the last
part of the period of inflaton
oscillations where the induced mass
in Eq.~\eq{effec-mass} can be
neglected. For small enough $\p$'s,
the potential $V(\p)$ can be
approximately taken to be quadratic
and this equation becomes of the
Emden type
\begin{equation}
\ddot\p+\frac{2}{t}\dot\p+
m_{3/2}^2\p=0.
\label{emden}
\end{equation}
The solution of this equation is
\begin{equation}
\p=\frac{A}{t}\sin m_{3/2}(t-t_0),
\end{equation}
where $A$ and $t_0$ are constants
depending on the initial conditions.
For $A<0$ and $0<m_{3/2}t_0\sim 1$,
the field initially (i.e. for $t
\lesssim m_{3/2}^{-1}\sim t_0$)
decreases approximately as
$-A\sin m_{3/2}t_0/t$. However, as
$t$ exceeds $m_{3/2}^{-1}\sim t_0$,
it is set in damped oscillations
about zero. On the other hand, for
$A<0$ and $0<m_{3/2}t_0\ll 1$, the
field initially decreases and
reaches zero at $t=t_0$. It then
passes to negative values and is
stabilized for a while (i.e. for
$t_0\lesssim t\lesssim
m_{3/2}^{-1}$) around $Am_{3/2}<0$.
It, finally, falls back into zero
and starts performing damped
oscillations about it. The former
behavior is encountered when $\p_f$
lies in the
lower part of the 2nd band, while the
latter when $\p_f$ reaches the upper
part of this band. We see that, in
the latter case, the field $\p$, as it
climbs up the potential towards the
maximum at $\p=-\p_{\rm max}$, `breaks'
and enters for a while in a plateau.
This makes it more difficult for $\p$
to reach this maximum. Thus, the upper
boundary of the 2nd band is pushed to
higher values.

\subsection{The inflationary perturbation
of the curvaton}
\label{subsect:perturbation}

\par
We are now ready to discuss the evolution
of the inflationary perturbation of $\p$
and estimate numerically the resulting
partial curvature perturbation $\zeta_\p$
at curvaton decay. For any value of $\l$
in the range $10^{-4}-10^{-3}$, we take
two values of $\p$ at the end of inflation:
$\p_f$ and $\p_f+\delta\p$ ($\delta\p=
H_{\rm infl}/ 2\pi$) both lying in the 3rd
band and follow numerically their evolution.
(Recall that $\delta\p$ is tiny compared
with the width of this band.) For both
these initial values, the field ends up
performing damped oscillations about the
PQ vacuum at $\p=-f_a$. When the amplitude
$\p_0$ of these oscillations is adequately
reduced, the oscillations become harmonic
with their amplitudes differing by an
amount $\delta\p_0$. The partial curvature
perturbation $\zeta_\p$ is then stabilized
to the limiting value $2\,\delta\p_0/3\p_0$,
as shown in section~\ref{sect:curvaton}.
This occurs well before the curvaton decay
in all the cases which we considered. We
find numerically the limiting $\zeta_\p$
for $10^{-4}\leq\l\leq 10^{-3}$
and any $\p_f$ in the 3rd band.

\par
It should be mentioned in passing that,
for a scalar field with a positive
power-law potential and a subdominant
energy density in a matter (radiation)
dominated universe, Eq.~\eq{field-eqn}
possesses \cite{ls99} an exact
scaling solution, which behaves as
a stable attractor provided that the
power is greater than 6 (10). In this
case, the field is eventually trapped
in the attractor and its energy
density decreases relative to the
dominant matter (radiation) energy
density as time elapses. Moreover,
any primordial perturbations in this
field are washed out since the field
falls in the same attractor no matter
what its initial conditions were.
Consequently, under these
circumstances, the field could not
play the role of a curvaton since the
limiting $\zeta_\p$ would come out
tiny. Fortunately, all the powers
involved in our PQ potential in
Eq.~\eq{PQ-pot} are less than or
equal to 6 and, thus, no attractor
type behavior is encountered with
this potential as our numerical
findings also clearly show.
Furthermore, if our full potential
$V(\p)$ in Eq.~\eq{full-pot} is
dominated by the variable induced
mass in Eq.~\eq{effec-mass}, new
scaling type solutions appear
\cite{prep}. However, the initial
perturbations are not washed out
in this case.

\FIGURE{\epsfig{file=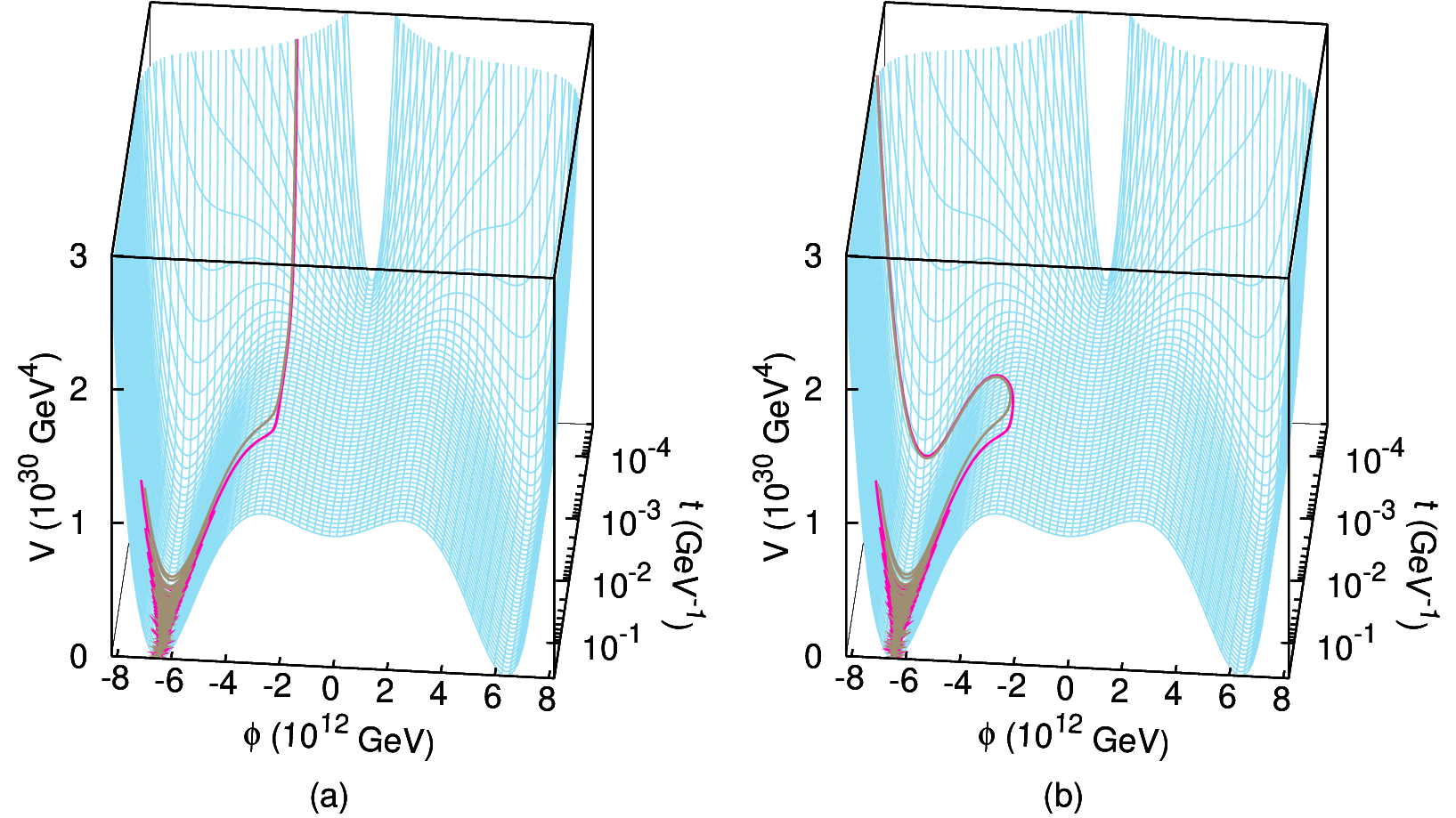,
width=8.426cm,angle=-90}
\caption{The evolution of the field
$\p$ for two $\p_f$'s differing by a
small amount and lying in the 3rd
band near its (a) lower, or (b) upper
boundary at $\l=10^{-4}$. The magenta
(brown) field paths correspond to
$\p_f$ being nearer (further from )
this boundary. The maximum at
$\p=-\p_{\rm max}$ acts as a
`tachyonic amplifier' of the
difference between the fields.}
\label{pot-tachyon}}

\par
As already indicated, the limiting
$\zeta_\p$ comes out generally small.
However, if $\p_f$ is near the upper
or lower boundary of the 3rd band,
$\zeta_\p$ can be enhanced
considerably. To see
this, let us consider two $\p_f$'s
slightly above (below) the lower
(upper) boundary of this band which
differ by a small amount $\delta\p$.
We take again $\l=10^{-4}$. The
relevant parts of the resulting
field paths are shown in
Fig.~\ref{pot-tachyon}a
(Fig.~\ref{pot-tachyon}b). The
magenta path corresponds to the
$\p_f$ which is nearer the boundary,
while the brown one to the $\p_f$
furthermost from the boundary. We
observe that the `magenta field' is
stabilized nearer the maximum at $\p
=-\p_{\rm max}$ than the `brown field'
after its first slow (its slow return
from its first slow) oscillation to
negative values. Then, due to the
flatness of the potential around the
maximum, the `magenta field' remains
near the maximum longer than the
`brown field'. Consequently, it falls
towards the PQ minimum at $\p=-f_a$
with a relative delay and, thus, the
gap $\delta\p$ between the fields
increases considerably. This leads
to a sizeable enhancement of the
disparity in the amplitudes of the
subsequent damped oscillations of
the two fields about the minimum.

\par
We see that the local maxima can
act as `tachyonic amplifiers' of
the curvaton perturbation which
originates from inflation. Note,
however, that the amplification
happens only if $\p_f$ is close
to the boundary of the band and
only once during the field
evolution, since, at subsequent
times, the field never approaches
the local maxima again. This
effect is crucial for the
viability of our model since
otherwise the limiting $\zeta_\p$
turns out too small. It should be
mentioned that
an amplification of perturbations
while the field is near a local
maximum of the potential was also
encountered in
Ref.~\cite {tachyonic}. However,
the phenomenon there was of a
quantum nature referring to the
quantum perturbations of the
inflaton in de Sitter space. In
our case, the effect is completely
classical and takes place well
after the termination of inflation.

\subsection{The curvaton decay}
\label{subsect:decay}

\par
We now turn to the discussion of
the curvaton decay and the
evaluation of the final curvature
perturbation. The curvaton decays
into a pair of Higgsinos via the
second coupling in the
superpotential of
Eq.~\eq{superpotential}. This same
coupling is also responsible for
the $\mu$-term. Of course, for this
decay to be kinematically possible,
we must make sure that the Higgsino
mass $\mu=\beta f_a^2/4m_P$ does
not exceed half of the curvaton mass
which is given by
\begin{equation}
m_{\p}^2=V_{\rm PQ}''(\p=\pm
f_a)=\frac{1}{3}\sqrt{|A|^2-12}
\left(|A|+\sqrt{|A|^2-12}\right)
m_{3/2}^2.
\end{equation}
This mass is $\l$-independent and
equals about $965~{\rm GeV}$ for
the input numbers used. The
curvaton decay time $t_{\p}=
\Gamma_{\p}^{-1}$, where
$\Gamma_{\p}$ is the curvaton
decay width given by
\begin{equation}
\Gamma_{\p}=\frac{\beta^2f_a^2}
{8\pi m_P^2}m_{\p}.
\end{equation}
We take $\mu=300~{\rm GeV}$, which
yields $\beta\simeq 0.697\l$ and
$t_\p\simeq 7.6~\l^{-1}\times
10^{13}~{\rm GeV}^{-1}$.

\par
It is important to ensure that the
coherently oscillating curvaton
field does not evaporate \cite{evap}
as a result of scattering with
particles in the thermal bath
before it decays into Higgsinos.
During the period between the onset
of field oscillations about a PQ
vacuum and roughly the electroweak
phase transition, the Higgs
superfields are in equilibrium with
the thermal bath and could knock the
PQ field out of the zero mode. The
dominant process is the
Higgsino-curvaton scattering through
the second coupling in the RHS of
Eq.~\eq{superpotential}. The cross
section is $\sigma\sim\beta^2/m_P^2$,
which yields the thermal decay rate
$\Gamma_{\rm th}\sim\beta^2T^3/m_P^2$.
Avoidance of evaporation then requires
that $\beta^2T^3/m_P^2\lesssim H$.
Before (after) reheating, $H\sim T^4/
m_PT_{\rm reh}^2$ \cite{reheat}
($H\sim T^2/m_P$) and this requirement
becomes $T\gtrsim\beta^2T_{\rm reh}^2
/m_P$ ($T\lesssim m_P/\beta^2$), which
is well-satisfied.

\FIGURE{\epsfig{file=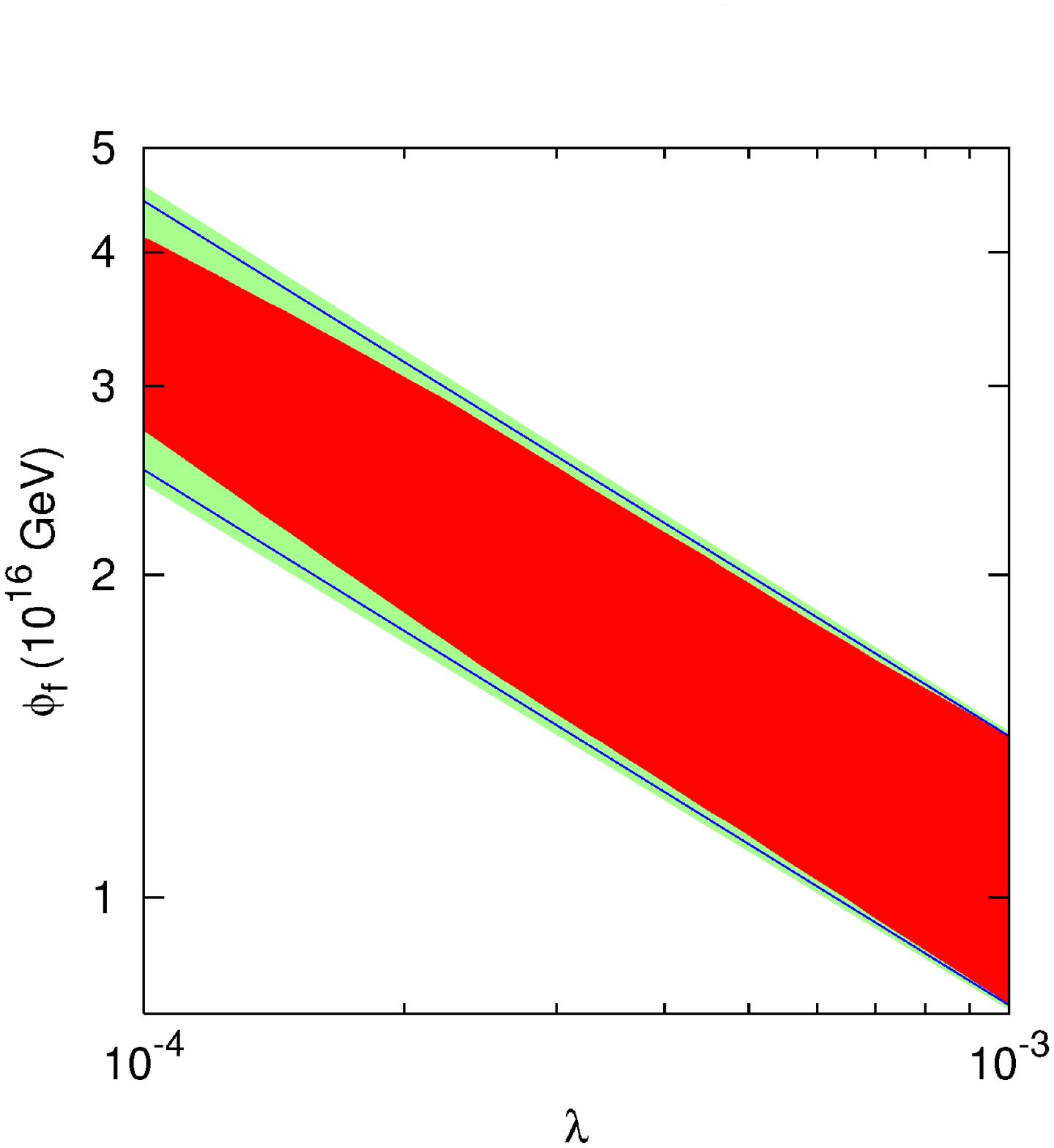,
width=12.04cm}
\caption{The 3rd band (all the
colored area) with the blue solution
lines on which the COBE constraint
$\zeta=5\times 10^{-5}$ is
satisfied. The green (red) area of
the band is allowed (excluded) by
the non-Gaussianity bound from the
WMAP satellite.}
\label{eps-zhet}}

\par
The energy density of the
oscillating curvaton at $t_\p$ is
evaluated numerically for each
$\p_f$ in the 3rd band and each
$\l$ in the range $10^{-4}-
10^{-3}$. From Eq.~\eq{f}, we can
then calculate $f_{\rm dec}$, the
fraction $f$ at curvaton decay.
The final curvature perturbation
$\zeta$ is estimated, in the
instantaneous decay approximation,
by multiplying this fraction by
the limiting $\zeta_\p$ (see
Eq.~\eq{fdec}). The result is put
equal to $5\times 10^{-5}$ as
required by COBE \cite{cobe}. We
find that there are two solutions
for each $\l$, one close to the
lower boundary of the band and
the other close to the upper
boundary. These solutions lie on
the two blue lines in
Fig.~\ref{eps-zhet}, where the
green and red areas together
constitute the 3rd band. It is
true that the solutions turn out
to be quite close to the
boundaries of the band. This
means that the `tachyonic
amplification' of the inflationary
perturbation in the curvaton
field, which occurs near a
maximum of the potential, plays a
crucial role. On the other hand,
the gap between the solutions and
the boundaries is not so tiny as
to imply that the solutions are
finely tuned. Moreover, this gap is
huge compared to the inflationary
perturbation $\delta\p\simeq 3.14
\times 10^{10}~{\rm GeV}$.

\par
We already stressed that, in the
curvaton scenario, significant
non-Gaussianity of the curvature
perturbation appears if the
curvaton decays before dominating
the energy density of the universe.
The recent CMBR data from the WMAP
satellite yield \cite{nongauss}
an upper bound on the possible
non-Gaussianity in the curvature
perturbation. This bound implies
\cite{luw02} that, at $95\%$ c.l.,
the fraction $f_{\rm dec}\gtrsim
9.33\times 10^{-3}$. The
corresponding allowed (excluded)
area of the 3rd band is colored
green (red) in Fig.~\ref{eps-zhet}.
We see that the (blue) solution
lines lie totally in the green
area and, thus, are fully
consistent with the negative
results of the WMAP satellite on
non-Gaussianity.

\par
As already explained, if $\p_f$
lies in a green band and near one
of its boundaries, the PQ field,
before starting its damped
oscillations about a PQ vacuum,
stays for a while near a maximum
of the potential. This `tachyonic
effect' is also important for the
fulfillment of the non-Gaussianity
constraint. Indeed, it causes a
delay in the commencement of the
damped oscillations of $\p$ and,
thus, leads to a larger $\rho_\p$
at curvaton decay. Therefore, the
fraction $f_{\rm dec}$ is
enhanced. As a consequence, the
non-Gaussianity bound can be
satisfied provided that $\p_f$ is
adequately close to the boundaries
of the band. This explains the
fact that the green area in
Fig.~\ref{eps-zhet} extends near
the boundaries of the band.
Needless to say that the
enhancement of $f_{\rm dec}$ also
facilitates the fulfillment of the
COBE constraint on the curvature
perturbation.

\par
In summary, we conclude that, in
the investigated scheme which
simultaneously solves the strong CP
and $\mu$ problems, the PQ field
can successfully and naturally act
as a curvaton with all the relevant
cosmological requirements satisfied.

\section{Conclusions}
\label{sect:conclusion}

\par
We considered a minimal extension of
MSSM which simultaneously solves the
strong CP and $\mu$ problems via a
PQ and a continuous R symmetry. This
model can be readily embedded in
simple and natural SUSY GUTs which
incorporate the hybrid inflationary
scenario and generate the observed
baryon asymmetry of the universe
through a primordial leptogenesis.
We, thus, took the model
supplemented with hybrid inflation
and leptogenesis, but without
committing ourselves to the specific
details of these scenarios.

\par
We examined whether, in this model,
the PQ field can play the role of
the curvaton generating the
primordial density perturbations
which are needed for explaining the
structure formation in the universe
and the observed anisotropies in
the CMBR. A crucial requirement is
that the PQ field is light during
the last $50-60$ inflationary
e-foldings so that it receives
a superhorizon spectrum of
perturbations from inflation.
This condition can be easily
satisfied in our model with global
SUSY, where the PQ potential is
almost flat, provided that, during
inflation, the value of the PQ field
is not too large. We must, though,
employ some mechanism which can
prevent the lifting of the flatness
of the potential by the SUGRA
corrections during inflation.

\par
We followed the evolution of the
PQ field after the end of inflation
under the assumption that the SUSY
breaking corrections to its
potential from the
finite energy density of the early
universe are somewhat small. We
found that the field, after an
initially slow evolution, is set in
damped oscillations about zero or
one of the PQ vacua, depending on
its value right after inflation.
More precisely, the values of the
PQ field at the end of inflation
can be classified into successive
bands which lead to the trivial or
the PQ vacua in turn. Needless to
say that only the bands leading to
one of the PQ vacua are useful for
our purposes.

\par
In our model, as it turns out, the
bulk of the CDM in the universe
consists of axions which are
produced at the QCD phase
transition. They carry an
isocurvature perturbation which is
uncorrelated with the total
curvature perturbation generated
by the PQ field. The rest of
CDM is made of LSPs (lightest
neutralinos) which originate from
the late decay of the primordial
gravitinos produced thermally at
reheating. These neutralinos as
well as the baryons, which are
assumed to come from a primordial
leptogenesis that took place at
reheating, acquire an isocurvature
perturbation fully correlated with
the curvature perturbation. So,
the overall isocurvature
perturbation has a mixed
correlation with the total
curvature perturbation. We found
that the presently available
restriction on such an
isocurvature perturbation from the
CMBR and other data excludes the
lowest lying band which leads to a
PQ vacuum. High order bands are also
excluded since they do not fulfill
the condition that the curvaton is
light during inflation. We are,
finally, left with only one allowed
band leading to a PQ vacuum. We,
thus, focus our attention to this
particular band.

\par
The inflationary perturbation of
the PQ field evolves as the cosmic
time elapses after the end of
inflation and, when the PQ field
is set in harmonic
oscillations about a PQ vacuum,
yields a stable perturbation in
the energy density of this field.
At the decay of the PQ field, the
perturbation is transferred to the
radiation dominated plasma
generating the total
curvature perturbation. We found
that the COBE constraint on the
total curvature perturbation can
be naturally fulfilled provided
that we are somewhat near the
boundaries of the band. An
important phenomenon, which
allows us to achieve this, is the
`tachyonic amplification' of the
original inflationary perturbation
in the PQ field as this field is
temporarily stabilized near a
local maximum of the potential.

\par
Finally, we imposed the bound on
the possible non-Gaussianity of the
total curvature perturbation from
the recent CMBR data obtained by
the WMAP satellite. We saw that,
although the bulk of the band is
excluded by this constraint, the
COBE requirement is still fully
consistent with it. The `tachyonic
effect', i.e. the fact that the
field hangs around a maximum of
the potential for a period of time
if its value right after inflation
is close to the boundaries of the
band, is crucial for having the
non-Gaussianity requirement
satisfied. The reason is that, due
to this effect, the commencement
of the damped field oscillations
is delayed and, thus, the
density fraction of the curvaton
is enhanced. This also facilitates
the achievement of the COBE
constraint on the curvature
perturbation.

\par
In conclusion, we have shown that,
in our scheme, which naturally and
simultaneously solves the $\mu$ and
strong CP problems, the PQ field
can successfully play the role of
the curvaton generating the total
curvature perturbation in the
universe in accord with the COBE
measurements. The bounds on the
isocurvature perturbation and
non-Gaussianity from the available
CMBR data can also be satisfied.

\bigskip

\acknowledgments
We would like to thank C. Gordon and
A. Lewis for communicating to us their
results on the bound from the
isocurvature perturbation prior to
publication. This work was supported
in part by the EU Fifth Framework
Networks `Supersymmetry and the Early
Universe' (HPRN-CT-2000-00152) and
`Across the Energy Frontier'
(HPRN-CT-2000-00148).

\end{document}